\ifdefined\pdfminorversion\pdfminorversion=7\fi
\documentclass[acmsmall,screen,nonacm]{acmart}
\setcopyright{none}
\authorsaddresses{}
\makeatletter
\let\hyxmp@parse@acmart\relax
\makeatother
\hypersetup{keeppdfinfo=true,pdflang={en}}
\AtBeginDocument{\hypersetup{
  pdfauthor={Ziyuan Li, Zihe Song, Shuxiao Xie, Mingfang Ji, Xiaofang Qi, Wei Yang, Tao Xie},
  pdfsubject={Validation reliability of generated GPU kernels},
  pdfkeywords={GPU kernels, differential testing, mutation testing, validation reliability, generated code},
  pdfcontactemail={taoxie@pku.edu.cn, yang\_wei@fudan.edu.cn},
  pdfcontactaddress={School of Computer Science and Engineering, Southeast University},
  pdfcontactcity={Nanjing},
  pdfcontactcountry={China}
}}

\usepackage{booktabs}
\usepackage{needspace}
\usepackage{flafter}
\usepackage{tabularx}
\usepackage{multirow}
\usepackage{makecell}
\usepackage{xspace}
\usepackage{listings}
\usepackage{algorithm}
\usepackage{algpseudocode}
\usepackage{enumitem}
\usepackage{subcaption}
\usepackage{tikz}
\usetikzlibrary{arrows.meta,positioning}
\usepackage[capitalize,noabbrev]{cleveref}
\newcommand{\ourtool}{\textsc{KernelMutate}\xspace}
\newcommand{\Mfull}{\textsc{M-full}\xspace}
\newcommand{\Mdir}{\textsc{M-dir}\xspace}
\newcommand{\pass}{\textsc{Pass}\xspace}
\newcommand{\fail}{\textsc{Fail}\xspace}
\newcommand{\inconclusive}{\textsc{Inconclusive}\xspace}
\newcommand{\rqanswer}[1]{\par\noindent\textbf{Answer.} #1\par}
\newcolumntype{Y}{>{\raggedright\arraybackslash}X}
\newcommand{\projecturl}{https://github.com/ZIYUANLI-star/KernelMutate}

\usepackage{xltabular}
\newcolumntype{Z}[1]{>{\hsize=#1\hsize\linewidth=\hsize}Y}
\newcolumntype{R}[1]{>{\hsize=#1\hsize\linewidth=\hsize\raggedleft\arraybackslash}X}
\title[KernelMutate: GPU Kernel Validation]{\ourtool: Mutation-Guided Validation of Generated GPU Kernels}
\author{Ziyuan Li}
\authornotemark[2]
\email{lziyuan32@gmail.com}
\affiliation{\institution{School of Computer Science and Engineering, Southeast University}\city{Nanjing}\country{China}}
\affiliation{\institution{Beijing Tongming Lake Information Technology Application Innovation Center}\city{Beijing}\country{China}}
\author{Zihe Song}
\authornotemark[2]
\email{Zihe.song@utdallas.edu}
\affiliation{\institution{The University of Texas at Dallas}\city{Richardson}\state{TX}\country{USA}}
\author{Shuxiao Xie}
\email{26113050150@m.fudan.edu.cn}
\affiliation{\institution{College of Computer Science and Artificial Intelligence, Fudan University}\city{Shanghai}\country{China}}
\affiliation{\institution{Beijing Tongming Lake Information Technology Application Innovation Center}\city{Beijing}\country{China}}
\author{Mingfang Ji}
\email{26113050055@m.fudan.edu.cn}
\affiliation{\institution{College of Computer Science and Artificial Intelligence, Fudan University}\city{Shanghai}\country{China}}
\affiliation{\institution{Beijing Tongming Lake Information Technology Application Innovation Center}\city{Beijing}\country{China}}
\author{Xiaofang Qi}
\email{xfqi@seu.edu.cn}
\affiliation{\institution{School of Computer Science and Engineering, Southeast University}\city{Nanjing}\country{China}}
\author{Wei Yang}
\authornote{Corresponding authors: Tao Xie and Wei Yang.}
\email{yang_wei@fudan.edu.cn}
\affiliation{\institution{College of Computer Science and Artificial Intelligence, Fudan University}\city{Shanghai}\country{China}}
\affiliation{\institution{Beijing Tongming Lake Information Technology Application Innovation Center}\city{Beijing}\country{China}}
\affiliation{\institution{Institute of Systems for Advanced Computing, Fudan University}\city{Shanghai}\country{China}}
\author{Tao Xie}
\authornotemark[1]
\email{taoxie@pku.edu.cn}
\affiliation{\institution{School of Computer Science and Engineering, Southeast University}\city{Nanjing}\country{China}}
\affiliation{\institution{Beijing Tongming Lake Information Technology Application Innovation Center}\city{Beijing}\country{China}}
\affiliation{\institution{School of Computer Science, Peking University}\city{Beijing}\country{China}}
\affiliation{\institution{College of Computer Science and Artificial Intelligence, Fudan University}\city{Shanghai}\country{China}}
\affiliation{\institution{Institute of Systems for Advanced Computing, Fudan University}\city{Shanghai}\country{China}}

\renewcommand{\shortauthors}{Ziyuan Li et al.}

\makeatletter
\g@addto@macro\@authornotes{%
  \footnotetext[2]{Equal contribution.}%
}
\renewcommand{\@mkauthors}{%
  \global\setbox\mktitle@bx=\vbox{%
    \unvbox\mktitle@bx
    \begingroup
    \centering
    \setlength{\parindent}{0pt}
    {\large
    Ziyuan Li\textsuperscript{1,2,\ensuremath{\dagger}}\quad
    Zihe Song\textsuperscript{3,\ensuremath{\dagger}}\quad
    Shuxiao Xie\textsuperscript{4,2}\quad
    Mingfang Ji\textsuperscript{4,2}\par
    \vspace{3pt}
    Xiaofang Qi\textsuperscript{1}\quad
    Wei Yang\textsuperscript{4,2,5,*}\quad
    Tao Xie\textsuperscript{1,2,6,4,5,*}\par}
    \vspace{7pt}
    {\small
    \textsuperscript{1}School of Computer Science and Engineering,\par
    Southeast University, Nanjing, China\par
    \textsuperscript{2}Beijing Tongming Lake Information Technology Application\par
    Innovation Center, Beijing, China\par
    \textsuperscript{3}The University of Texas at Dallas, Richardson, TX, USA\par
    \textsuperscript{4}College of Computer Science and Artificial Intelligence,\par
    Fudan University, Shanghai, China\par
    \textsuperscript{5}Institute of Systems for Advanced Computing,\par
    Fudan University, Shanghai, China\par
    \textsuperscript{6}School of Computer Science, Peking University, Beijing, China\par}
    \vspace{6pt}
    {\footnotesize
    \href{mailto:lziyuan32@gmail.com}{\nolinkurl{lziyuan32@gmail.com}},
    \href{mailto:Zihe.song@utdallas.edu}{\nolinkurl{Zihe.song@utdallas.edu}}\par
    \href{mailto:26113050150@m.fudan.edu.cn}{\nolinkurl{26113050150@m.fudan.edu.cn}},
    \href{mailto:26113050055@m.fudan.edu.cn}{\nolinkurl{26113050055@m.fudan.edu.cn}}\par
    \href{mailto:xfqi@seu.edu.cn}{\nolinkurl{xfqi@seu.edu.cn}},
    \href{mailto:yang_wei@fudan.edu.cn}{\nolinkurl{yang_wei@fudan.edu.cn}},
    \href{mailto:taoxie@pku.edu.cn}{\nolinkurl{taoxie@pku.edu.cn}}\par}
    \endgroup
    \vspace{10pt}%
  }%
}
\makeatother

\begin{document}
\begin{abstract}
Generated GPU implementations are commonly accepted after a few differential tests, although replacement obligations may remain unchecked.
We present \ourtool, an empirical study of validation reliability using controlled faults, released implementations, and replacement tests.
A five-draw anchor accepts 548 of 838 CUDA candidates.
The full suite finds native FP32 witnesses for 136 accepted candidates and dtype-only findings for another 362, mostly low-precision execution limitations without established publisher guarantees.
Source-guided testing identifies the full suite's 692 candidates at 26.1\% lower charged cost, but yields only dtype evidence for 16 accepted candidates with native witnesses in the full suite.
Independently observed context and boundary findings cover the same candidate set.
Task-held-out template matching reveals differences in ranking and observation coverage: global top-eight recommendations match more probes than class-conditioned lists (51 versus 48 of 75), while conditional analyses restrict the apparent advantage at smaller limits.
These template-matching results are offline measurements, not deployed-suite gains.
Independent controls and replacement cases expose numerical-policy, binding, and custom-execution boundaries.
Validation therefore needs to report witness domains and observation coverage alongside candidate yield and cost.
\end{abstract}

\ccsdesc[500]{Software and its engineering~Software testing and debugging}
\ccsdesc[300]{Software and its engineering~Empirical software validation}

\keywords{GPU kernels, differential testing, mutation testing, validation reliability, generated code}
\maketitle

\section{Introduction}
\label{sec:intro}
\begin{figure}[htbp]
\centering
\includegraphics[width=\linewidth]{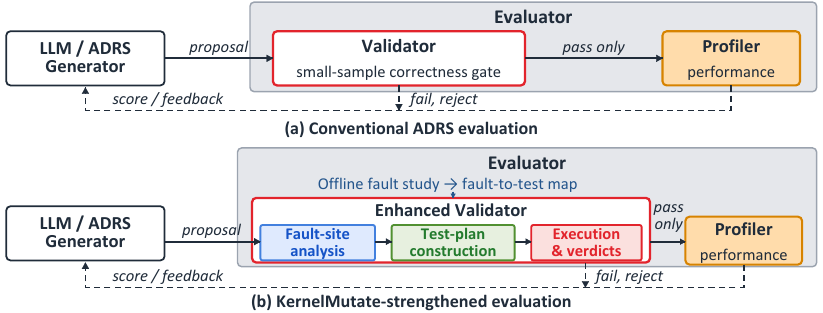}
\caption{Conceptual placement of \ourtool in a kernel-generation workflow, using an offline fault-to-test map.
Only accepted candidates reach performance measurement.
Inconclusive outcomes (not shown) remain distinct from failures.}
\Description{Two panels place an ordinary or source-guided validator between a candidate generator and a profiler.
Dashed arrows return validation and performance feedback to the generator.}
\label{fig:adrs}
\end{figure}

Kernel validators are essential for assessing whether generated GPU implementations preserve the behavior of reference computations.
Kernel-generation systems compare candidate implementations with references on sampled inputs before measuring performance~\cite{ouyang2025kernelbench,cudaL1,aicudaengineer}.
In AI-Driven Research for Systems (ADRS) workflows, accepted candidates receive performance feedback that guides further generation (Figure~\ref{fig:adrs})~\cite{cheng2025letbarbarians}.
When validation misses a discrepancy, the resulting performance feedback can favor an implementation that changes the intended computation.
Conversely, passing numerical checks does not establish replacement efficiency: DSL kernels can remain substantially slower than optimized library operators~\cite{li2026correctbutslow}.

Improvements to kernel validators address both the range of test conditions and the reliability of correctness checks.
Tests vary initialization, configurations, and input distributions to compare candidate and reference behavior across different conditions~\cite{robustkbench,kernelbenchverified}.
Studies of GPU numerical bugs and tile-program code generation further identify input values, shapes, dtypes, and backend behavior as important failure-triggering conditions~\cite{rathnasuriya2025gpunumerical,rathnasuriya2026tilebugs}.
Interpreting the observed discrepancies requires criteria that distinguish implementation errors from acceptable numerical variation.
Tolerance calibration sets comparison thresholds according to the operator and dtype~\cite{gpuemutolerance}.
Beyond numerical tolerances, contract-based checks examine properties such as preservation of non-finite values.
Known-correct implementations help assess false alarms~\cite{contractgrade}, while mutation analysis uses controlled faults to assess the ability of test cases to detect errors~\cite{measuringchecker}.

The remaining difficulty is choosing informative test cases within a limited budget and interpreting the resulting evidence for replacement.
Repeated samples from one distribution can leave relevant behavior untested.
For example, tests with only nonnegative inputs cannot expose an in-place ReLU that overwrites negative input values while the reference preserves its inputs.
Covering more conditions consumes additional testing resources, so validation needs a way to prioritize test cases.
Test selection alone does not determine what a failure or a pass establishes.
A numerical failure under the native FP32 contract differs from a candidate's rejection of an FP16 input.
The FP16 rejection concerns support for that dtype and does not by itself establish incorrect FP32 behavior.
A passing comparison shows that the executed code meets the comparison criteria on the tested inputs.

We present \ourtool, a study of validation reliability that combines fault-informed test selection with checks on the scope of validation evidence.
Controlled mutations identify conditions that expose implementation differences.
We organize the observed associations between fault classes and test conditions into a fault-to-test map.
\Mdir uses the map and candidate source sites to prioritize test templates, then fills the remaining budget with general tests.
\Mfull provides a fixed suite of value and execution-context tests for comparison.
To interpret the outcomes, we record each failure under the study's validation rules as a \emph{witness}, retaining the failure stage and tested domain.
The records distinguish failures under the native FP32 contract from failures involving additional FP16 or BF16 inputs.
A separate replacement study examines reference bindings and actual custom execution in the tested contexts.

Our experiments show that broader testing finds witnesses missed by repeated random sampling, while \Mdir incurs less charged testing cost than \Mfull.
Across 838 released CUDA candidates, \Mdir and \Mfull find witnesses for the same 692 candidates, compared with 179 under up to 32 IID draws.
Combining the recorded findings from independent boundary and dtype tests also covers the same candidate set.
Under the study's cost-accounting rules, \Mdir incurs 26.1\% lower testing cost than \Mfull.
The comparison between \Mdir and \Mfull evaluates complete strategies and does not isolate the effect of source guidance.

Interpreting the candidate counts requires distinguishing the tested domains.
Of 548 candidates accepted by five random draws, \Mfull finds native FP32 witnesses for 136 and only additional-dtype witnesses for another 362.
Most additional-dtype witnesses concern low-precision rejections without established publisher guarantees of support.
For 16 of these accepted candidates, \Mfull finds native FP32 witnesses while \Mdir finds only additional-dtype witnesses.
A separate replacement check finds a shape error in a custom implementation that the passing wrapper tests did not execute.

This work contributes empirical evidence, a fault-informed testing method, and an analysis of what validation outcomes establish:
\begin{itemize}
\item \textbf{Evidence of validation gaps.}
We study 838 released CUDA candidates and 1,646 controlled probes to characterize native FP32 failures, additional-dtype limitations, and conditions that expose surviving probes.
\item \textbf{Fault-informed test construction.}
We develop an empirical fault-to-test map and budgeted source-guided plans, evaluated through task-held-out recommendations and comparisons of witness yield, tested domain, and cost.
\item \textbf{Analysis of validation outcomes.}
We use independent controls and replacement tests to examine how numerical policies, reference bindings, and actual custom execution affect validation outcomes.
\end{itemize}

We release the code and replication materials at \expandafter\url\expandafter{\projecturl}.
Appendix~\ref{sec:appendix-evidence} provides the extended evidence tables, comparison protocols, and witness analyses.

\section{Background and Motivation}
\label{sec:background}
Our \emph{validation contract} records the candidate entry point, paired reference, input domain, corresponding state, execution mode, and output policy. A generated implementation may replace a PyTorch~\cite{paszke2019pytorch} operation, fused block, or model; KernelBench supplies executable references and input generators~\cite{ouyang2025kernelbench}. The output policy permits specified differences, including floating-point variation from precision and evaluation order~\cite[Chs.~2 and~4]{higham2002accuracy}. Output comparisons require a valid reference and corresponding state; sampled agreement supports only the tested context.

A \emph{native FP32 witness} records a scored failure under the study's FP32 contract; an \emph{additional-dtype witness} concerns requested FP16 or BF16. These labels identify the requested dtype, not the failure cause (\cref{sec:verdicts}). Additional-dtype findings need not violate original publisher specifications. C3-23530's LeNet-5 agrees on tested FP32 inputs but rejects Half tensors before custom CUDA launch (\cref{tab:dtype-cases}). Further FP32 draws cannot reveal that rejection.

Execution also limits the evidence: C5-01485 defines custom CUDA, yet its published \texttt{forward()} only calls PyTorch layers (\cref{sec:replacement}). Wrapper agreement cannot validate the dormant function.

Even when the intended code executes under a valid contract, particular values or precision may be needed to expose discrepancies, motivating targeted tests.

\section{Controlled Fault Probes}
\label{sec:study}
\begin{figure}[htbp]
\centering
\includegraphics[width=\linewidth]{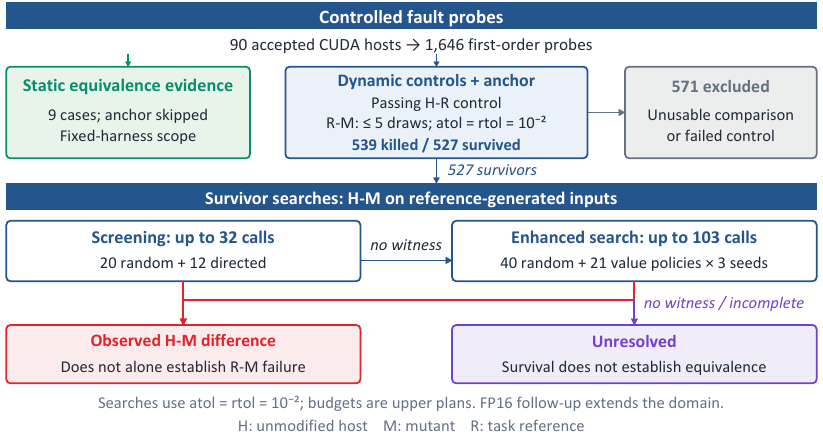}
\caption{Controlled-probe analysis. Nine cases with harness-scoped equivalence evidence bypass the anchor; dynamic records distinguish exclusions, anchor kills, and survivors eligible for further search. Budgets denote planned calls.}
\Description{Construction yields 1,646 probes. Nine cases have equivalence evidence under the fixed harness and skip the anchor. The dynamic branch records 571 excluded comparisons, 539 anchor kills, and 527 tested survivors. Survivors enter screening and enhanced search with plan limits of 32 and 103 calls. Search outcomes distinguish witnessed differences from unresolved probes. Additional FP16 witnesses extend the observed domain.}
\label{fig:study-workflow}
\end{figure}

We use controlled mutations to identify conditions that expose implementation differences and to construct an empirical prior for test selection~\cite{demillo1978,jiaharman2011}. Figure~\ref{fig:study-workflow} outlines probe construction, validity checks, and evidence classification.

\paragraph{Probe construction.}
C1 contains 1,646 first-order probes from 90 previously accepted CUDA hosts, selecting the highest recorded speedup among available turns for each selected KernelBench L1/L2 task. Each probe applies one source rewrite at a Python AST or CUDA-pattern site, retaining at most three per host--operator pair. The 16 operators span arithmetic, GPU parallelism, ML numerical behavior, and broadcast/layout assumptions. Appendix~\ref{sec:app-mutations} lists operators, transformations, and observed counts. Counts reflect this construction, not natural-fault frequencies.

\paragraph{Comparison and evidence.}
Dynamic evidence requires a usable comparison and a passing control for the unmodified host against the paired reference. The anchor compares mutant with reference on up to five FP32 draws at $\mathrm{atol}=\mathrm{rtol}=10^{-2}$. Host controls establish agreement on their recorded inputs; they do not validate every input used by the survivor searches.

Equivalence is scoped to the fixed harness~\cite{madeyski2013overcoming}. Each of the nine screened cases changes an integer used only by candidate \texttt{get\_inputs}, leaving CUDA, initialization, forward computation, and state unchanged. The harness calls reference generators instead, so compared behavior is unaffected (Appendix~\ref{sec:app-mutations}). These cases bypass dynamic mutant testing; finite survival alone does not establish equivalence.

Two searches seek inputs that distinguish dynamic anchor survivors from their unmodified hosts. Screening plans up to 32 calls (20 random + 12 directed). Probes without a witness are eligible for enhanced search with up to 103 calls (40 random + 21 value policies $\times$ three seeds). Both searches use task-reference inputs and $10^{-2}$ absolute/relative tolerances. Approximate agreement is not transitive: a host-relative witness need not be a reference-relative failure. Two additional FP16 witnesses extend the observed domain. Search probes without witnesses remain unresolved; timeouts and incomplete plans are recorded separately. Section~\ref{sec:rq4} gives the complete accounting.

\paragraph{Fault-to-test prior.}
Candidate testing uses one fixed screening-derived map with 607 template-supporting probes. Its singleton \texttt{F-IDX} association with the last-element boundary template lacks admissible support: the recorded rejection reason is nondeterminism. The template also has support from \texttt{F-BOUND} (8/9) and \texttt{F-CONST} (3/5), so it cannot be attributed solely to that association. The artifact binds the map and candidate plans; \cref{sec:rq4} measures the related witness dependence.

\paragraph{Task-held-out recommendation analysis.}
We separately measure transfer of confirmed fault conditions by five-fold cross-fitting over whole tasks, whose probes share a host and reference. The training roster contains 539 anchor kills and 68 confirmed screening witnesses; one lacks a template identity, leaving 606 probes with usable template evidence. Evaluation includes all 75 primary witnessed survivors, with seven enhanced-search cases used only for evaluation. The two additional FP16 cases remain outside this population.

Training-fold records rank templates by fault class or globally after pooling classes, using \cref{eq:selection-score}; ties use training mean time, then template identity. We compare caps from one to eight without filling exhausted lists. Coverage requires an identified recorded witness; unexecuted or inconclusive templates remain unknown. Probe classes are known from construction, unlike the multiple source-derived hypotheses for natural candidates. This measures template matching across seeds, without candidate replay.

\section{Fault-Guided Candidate Validation}
\label{sec:method}
\begin{figure}[htbp]
\centering
\includegraphics[width=\linewidth]{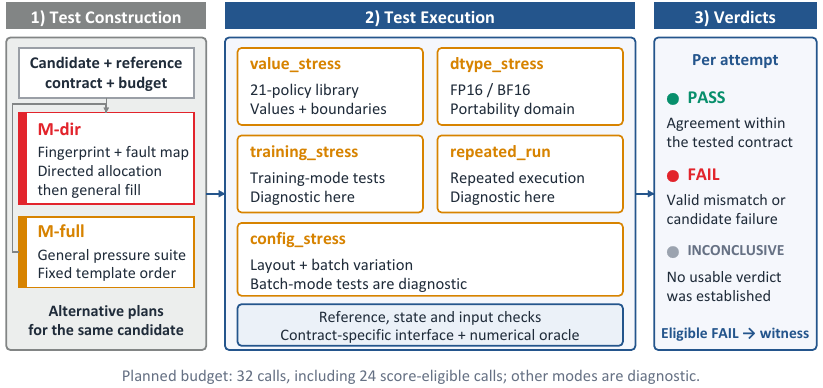}
\caption{Evaluated plans across five testing dimensions. \Mdir uses source sites and the fault-to-test map; \Mfull follows a fixed sequence. Validation stages determine verdicts; only score-eligible failures contribute discoveries. The plans also include unscored diagnostics.}
\Description{Two alternative plans select cases from value, dtype, training-mode, repeated-execution, and configuration tests. Validation attempts produce PASS, FAIL, or INCONCLUSIVE; eligible failures contribute candidate discoveries.}
\label{fig:tool-workflow}
\end{figure}

Given a candidate, paired reference, validation contract, and invocation budget $B$, \ourtool constructs directed (\Mdir) or full (\Mfull) plans (Figure~\ref{fig:tool-workflow}) and records their verdicts.

\subsection{Source Fingerprints and an Empirical Prior}
Source fingerprints identify eligible fault classes, excluding sites matching prior-equivalence patterns. A class requires both a remaining site and an effective map template. Site counts determine inclusion, not weights or budget shares; matches establish neither a defect nor execution of matched code.

The map ranks templates by conditional exposure rates. A \emph{template} comprises policy, mode, and parameters; a \emph{case} adds a seed. Let $W_f$ contain class-$f$ probes with identified witness templates in the map's input records (\cref{sec:study}). For template $t$, $E_{f,t}\subseteq W_f$ contains probes with a conclusive execution of $t$, excluding inconclusive and invalid-input outcomes. $K_{f,t}\subseteq E_{f,t}$ contains those with a witness under the stage's comparison: mutant--reference for anchor kills, mutant--host for screening differences. Each probe counts once per set across seeds. For nonempty $E_{f,t}$,
\begin{equation}
q(f,t)=|K_{f,t}|/|E_{f,t}|.\label{eq:selection-score}
\end{equation}
Class $f$ retains templates with $|K_{f,t}|>0$, storing four-decimal scores and both counts. Counts reveal sparse support without correcting selection effects; the ranking is not a natural-candidate defect probability.

\subsection{Constructing and Executing a Plan}
\Mdir combines directed selection and general tests within $B$ planned calls. Represented classes are prioritized by their best map score; their descending-score template lists are traversed round-robin, deduplicating templates. Ordinary cases cost one call; repeated cases cost their specified repetitions. A heuristic caps directed selection at $\lceil0.7B\rceil$ calls. At this cap or list exhaustion, unused templates are appended from a fixed general sequence covering IID/structured values, boundaries, dtype, repetition, and batch configurations. Cases exceeding a stage's remaining budget are skipped while selection continues. Residual capacity cycles through the general sequence. The experimental plans specify every case's order and seed, including repeated selections of a template.

General tests also permit construction without source matches or effective templates; neither equal class allocation nor optimal selection is guaranteed.

\Mfull follows a fixed sequence: 21 value policies covering magnitudes, sign patterns, sparsity, boundaries, and reduction-sensitive inputs, followed by dtype and execution-context cases.

Plans and score eligibility appear in \cref{tab:budgets}; execution and completion rules appear in \cref{sec:setup}, and cost accounting in \cref{sec:rq2}. The comparison evaluates complete constructions and execution coverage, not simply reordered tests.

\subsection{Validity and Verdicts}
\label{sec:verdicts}
The main-study oracle uses differential testing to compare observable behavior~\cite{mckeeman1998}. With matched inputs and state, it compares returned values, input side effects, and post-call module state. Nested structure and all corresponding leaves must agree, including tensor shape, dtype, device, and layout type; stride and aliasing equality are not required. NaN positions must match, with matching NaNs accepted; infinity positions and signs must match. Integer and Boolean elements require exact equality.

Finite floating-point reference and candidate elements $r_i,c_i$ must satisfy $|r_i-c_i|\leq\mathrm{atol}+\mathrm{rtol}|c_i|$, using \texttt{torch.isclose(reference, candidate)}~\cite{pytorch212isclose}. Absolute tolerance handles near-zero values; relative tolerance scales with candidate magnitude. We preset $\mathrm{atol}=\mathrm{rtol}=10^{-4}$ for FP32 and $10^{-2}$ for FP16/BF16 as acceptance policies, not universal numerical error bounds.

Verdicts follow the stage reached and the evidence available at that stage. Reference import/setup failures yield \inconclusive. Candidate import failures (after successful reference import) and candidate model setup failures yield \fail. Model setup includes class lookup, construction, device/dtype conversion, and mode setting. These \fail verdicts require neither reference forward execution nor completed state binding; the invocation-stage resource exclusion does not apply. Failures in validator state binding, snapshots, or input preparation yield \inconclusive, as do parent timeouts. During invocation, reference failures or recognized memory resource errors yield \inconclusive; other candidate exceptions after reference success, including compilation within \texttt{forward}, yield \fail. Valid comparisons yield \fail for numerical/interface mismatches and \pass for agreement.

A strategy counts each candidate once upon any selected, score-eligible \fail. Records retain failure stage, case/source identities, policy, seed, mode, parameters, and available timing/error evidence. Within a strategy, native FP32 evidence takes precedence; dtype-only means additional-dtype failures without an observed native witness (\cref{tab:witness-domains}). Passing observations concern only exercised inputs and callable; replacement tests check custom execution separately (\cref{sec:replacement}).

\section{Evaluation}
\label{sec:evaluation}
\label{sec:eval}
We study the reliability of small-sample acceptance, the benefit and cost of stronger tests, and the interpretation of their outcomes through five research questions:
\begin{enumerate}[label=\textbf{RQ\arabic*:},leftmargin=*]
\item What native and additional-dtype failures remain among naturally generated candidates accepted by a small suite?
\item How do input strategies compare in witness yield and cost, and how much does numerical tolerance explain?
\item How do validators behave on independently labeled bugs and controls?
\item Which controlled faults escape the anchor, and how well do their test conditions transfer across tasks?
\item Which suite components retain observed discoveries, and why does dtype require explicit testing?
\end{enumerate}
A complementary replacement study examines network integration, reference binding, and custom execution.

\subsection{Study Design}
\label{sec:setup}
\paragraph{Subjects and selection.}
We evaluate 838 collected CUDA implementations from C2, C3, and C5 across 238 task identifiers~\cite{cudaL1,aicudaengineer,kernelbenchsamples}. Eligibility requires nonempty source, a resolvable KernelBench task, and static evidence of custom CUDA. Whitespace-normalized source hashes remove within-corpus duplicates. C2 includes all 85 eligible accepted records. C3 contributes 227 candidates, selecting the lexicographically smallest full stable source ID among each task's eligible accepted records. C5 supplies 526 eligible candidates accepted in its published baseline-evaluation pool. Collection acceptance labels are distinct from B1u acceptance.

C1, C4, and C6 have distinct roles and are excluded from natural-candidate denominators (\cref{tab:corpora}). C1 informs test construction; C4 is inventory only. C6 retains third-party bug/control labels~\cite{gpuemu} and compares outputs to its supplied FP64 reference via \texttt{run(dict)}.

\begin{table}[htbp]
\caption{Corpus inventory and roles. Counts denote implementations or probes, not necessarily distinct tasks.}
\label{tab:corpora}
\small
\begin{tabularx}{\linewidth}{@{}lp{25mm}lrY@{}}
\toprule
ID & Collection & Backend & Count & Experimental role \\
\midrule
C1 & KernelBench L1/L2 mutations & CUDA & 1,646 & Controlled fault conditions and evidence for test construction \\
C2 & CUDA-L1 & CUDA & 85 & Natural candidates and replacement study \\
C3 & AI-CUDA-Engineer & CUDA & 227 & Natural candidates and replacement study \\
C4 & TritonBench-G & Triton & 184 & Inventory only: paired references not established \\
C5 & KernelBench-samples & CUDA & 526 & Natural candidates and replacement study \\
C6 & gpuemu & \makecell[l]{Triton/CUDA,\\CPU} & 26 & Third-party labels: ten seeded bugs and 16 controls \\
\bottomrule
\end{tabularx}
\end{table}

Candidates are identified by corpus, task, and stable source ID; different implementations of a task remain separate. We report comparisons and intervals for the 753-candidate C3+C5 panel alongside the full 838-candidate population.

C2, C3, and C5 overlap with C1 on 38, 90, and 85 tasks, covering 38, 90, and 316 candidates, respectively. Task-held-out cross-fitting evaluates recommendations; natural-candidate experiments evaluate complete strategies on the collected population. The sample is nonrandom; corpus membership does not establish source disjointness.

\paragraph{Comparison domains and validity.}
Main observations concern the specified callable and its validation stages under the study contract. Contracts fix recorded input shapes and adopt unrestricted finite values and determinism as study defaults, not independently audited publisher specifications. Native FP32 witnesses concern this contract; additional FP16/BF16 findings concern replacement support. Low-precision obligations remain unestablished for C2/C3; C5's recovered public evaluation setting is FP32. B1u, B3--B6, \Mfull, and \Mdir share the numerical/interface oracle in \cref{sec:verdicts}. Output comparisons require usable reference and state bindings; candidate import/setup failures follow the separate stage rules there. Replacement tests separately check custom execution.

\paragraph{Strategies and budgets.}
The strategy panel separates additional sampling, value and context variation, and complete fault-informed constructions (\cref{tab:validators}). B7--B9 retain the external policies' inputs and decision rules~\cite{robustkbench,kernelbenchx}; B10 checks dynamic CUDA properties~\cite{nvidiasanitizer}. B12 applies fixed tolerance to the full-suite schedule on the selected cohort in \cref{sec:rq2}.

\begin{table}[htbp]
\caption{Validator strategies and roles. IDs remain fixed across studies; interfaces depend on the corpus.}
\label{tab:validators}
\small
\begin{tabularx}{\linewidth}{@{}lp{34mm}Y@{}}
\toprule
ID & Strategy & Input construction and experimental role \\
\midrule
B1 & Default-style anchor & Five IID inputs; fixed $\mathrm{atol}=\mathrm{rtol}=10^{-2}$ \\
B1u & Dtype-aware anchor & Five IID inputs; FP32 tolerance $10^{-4}$, FP16/BF16 $10^{-2}$ \\
B3 & IID-32 & Unchanged input distribution and execution context; 32 seeds \\
B4 & Value-diversified & Eight value distributions $\times$ four seeds \\
B5 & Boundary values & Eight boundary policies $\times$ four seeds: zeros, extreme magnitudes, and boundary elements \\
B6 & Composite context & Eight dtype cases (FP16/BF16 $\times$ four seeds), eight training cases (IID/near-zero $\times$ four), and 16 batch cases ($\{1,4,16,64\}\times$ four); no fault prior \\
B7 & robust-kbench-style & External strategy port; its supported inputs and oracle \\
B8 & KernelBenchX-style & External strategy port; its supported inputs and oracle \\
B9 & Differential fuzzing & Seeded FP64 CPU-reference port; operator- and dtype-calibrated tolerances \\
B10 & Compute Sanitizer & Dynamic CUDA error/hazard checking \\
B11 & \Mdir & Code fingerprints and fault-informed construction; nominal $\approx70\%$ directed allocation with general budget fill \\
B12 & M-fixed & Full-suite input schedule with fixed $10^{-2}$ absolute/relative tolerance \\
B13 & \Mfull & 21 value policies plus dtype, training, repeated-execution, and configuration tests; 32-call limit \\
\bottomrule
\end{tabularx}
\end{table}

Plans distinguish logical cases, nominal invocations, and score-eligible invocations (\cref{tab:budgets}). Invocations count candidate calls, not CUDA launches; repeated cases consume multiple calls. Non-anchor controlled plans allow at most 32 calls, with at most 23 directed calls for \Mdir. Training, batch-mode, and repeated-mode diagnostics remain unscored. Main runs stop at the first \fail per candidate--strategy pair or after five consecutive timeouts; each test-case worker has a 180-second parent-process timeout. Completion runs attempt eligible cases omitted by these guards within the same plan, with failure stopping disabled. Reference/validator preparation failures and timeouts remain inconclusive. Each candidate--strategy--case contributes at most one selected observation; confirmations do not enlarge denominators.

\begin{table}[htbp]
\caption{Per-candidate plans for 838 natural candidates. Logical cases, nominal invocations, and score-eligible invocations are counted separately.}
\label{tab:budgets}
\small
\begin{tabularx}{\linewidth}{@{}lrrrYY@{}}
\toprule
Strategy & \makecell[r]{Logical\\cases} & \makecell[r]{Nominal\\calls} & \makecell[r]{Eligible\\calls} & Eligible composition & Other planned calls \\
\midrule
B1u & 5 & 5 & 5 & Five IID & None \\
B3 & 32 & 32 & 32 & 32 IID & None \\
B4 & 32 & 32 & 32 & 32 value & None \\
B5 & 32 & 32 & 32 & 32 boundary & None \\
B6 & 32 & 32 & 8 & Eight dtype & Eight training; 16 batch \\
\Mfull & 31 & 32 & 24 & 21 value; two dtype; one layout & Two training; four batch; two repeat \\
\Mdir & 30 & 32 & 24 & 20 value; four dtype & Four batch; four repeat \\
\bottomrule
\end{tabularx}
\end{table}

\Needspace{8\baselineskip}
\paragraph{Measures and inference.}
For natural candidates, witness yield is $|D_s|/N$: candidates with a score-eligible failure divided by the full population, including inconclusive candidates. It is not defect recall. First-witness time accumulates recorded calls through the first scored failure, including preceding unscored calls; reported medians condition on $D_s$ and use the upper middle observation for even counts. Cost separately aggregates charged testing wall time per 1,000 targets under the scopes in \cref{tab:cost}.

C3+C5 yield-difference intervals use paired task-cluster bootstrap resampling: tasks are drawn with replacement, retaining their implementations and paired strategy outcomes (10,000 resamples, seed 20260819). The intervals measure resampling uncertainty, not execution variability, and do not correct nonrandom selection.

The main and replacement studies use A800 (PyTorch 2.1.2, CUDA 12.1) and RTX 4090 (PyTorch 2.5.1, CUDA 12.4), respectively; B9 uses CPU. Cross-platform timings are not pooled. The main-study records do not establish global TF32, autocast, or reduced-precision-reduction settings; replacement settings are stated separately.

OpenAI Codex assisted implementation of CPU scripts for witness aggregation, cost accounting, cohort selection, and offline rankings. Outputs were checked against recorded identities, hashes, and separate aggregation implementations.

\subsection{RQ1: Evidence After Small-Sample Acceptance}
\label{sec:rq1}
All 548 B1u-accepted candidates have exactly five valid \pass observations and no inconclusive outcome. Of 838 candidates, B1u rejects 177 and leaves 113 undecidable. Acceptance requires at least one valid \pass and no \fail; all-inconclusive candidates are not accepted.

We classify accepted candidates using shared-oracle evidence pooled from B3--B6, \Mfull, and \Mdir (\cref{tab:accepted}). Native FP32 evidence takes precedence; dtype-only means additional-dtype evidence without a pooled native witness. Remaining candidates with B7--B10 findings are external-only; the rest have no qualifying finding. Precedence yields disjoint reporting categories despite overlapping witness sets.

\begin{table}[htbp]
\caption{Findings pooled across strategies for 548 B1u-accepted candidates. Percentages use 548; dtype-only findings are additional-domain evidence.}
\label{tab:accepted}
\small
\begin{tabularx}{\linewidth}{@{}Yrrrr@{}}
\toprule
Evidence category & C2 & C3 & C5 & Total (\%) \\
\midrule
Native FP32/value: B3--B5, \Mfull, \Mdir & 9 & 24 & 103 & 136 (24.8) \\
Dtype only: B6, \Mfull, \Mdir & 4 & 47 & 311 & 362 (66.1) \\
External validators only: B7--B10 & 8 & 0 & 4 & 12 (2.2) \\
No qualifying finding & 33 & 0 & 5 & 38 (6.9) \\
\midrule
B1u-accepted candidates & 54 & 71 & 423 & 548 (100.0) \\
\bottomrule
\end{tabularx}
\end{table}

Native symptoms include finite-value and NaN/Inf-position differences in outputs or side effects, rather than source-level defect categories (Appendix~\ref{sec:app-native}).

Additional-dtype findings mostly concern execution support. Among the 362 dtype-only candidates, 359 fail to execute on low-precision inputs while the reference succeeds; three C2 candidates return output discrepancies.

The disjoint native and dtype-only categories give 498/548 (90.9\%) with shared-oracle witnesses. Ten B9 and two sanitizer-only findings raise this count to 510/548 (93.1\%) across mixed rules. Neither fraction estimates native-defect prevalence; the 38 residual candidates are not certified correct.

\paragraph{Interpreting native witnesses.}
Native witnesses expose different contract divergences. In C2 L1-P19, \texttt{relu\_} mutates negative inputs while the reference's \texttt{relu} preserves them, a side effect invisible on nonnegative draws. In C2 L1-P37, denormals expose differing infinity positions: the candidate adds $10^{-12}$ before reciprocal square root, whereas the reference divides by an unregularized norm. The recorded size selects PyTorch fallback. These cases distinguish input-side-effect semantics from numerical conventions; neither establishes an executed custom-CUDA defect. Appendix~\ref{sec:app-native} gives identities and error evidence.
\rqanswer{Small-sample acceptance leaves both native discrepancies and additional-dtype limitations unchecked. The latter concern replacement support; a violation of the original specification additionally requires an established obligation.}

\subsection{RQ2: Effectiveness, Cost, and Numerical Strictness}
\label{sec:rq2}
\paragraph{Discovery under shared validation rules.}
On all 838 candidates, IID-32 yields 179 detections versus 177 for the five-draw anchor: three new detections and one not reproduced, a net gain of two (\cref{tab:main-results}). The runs use different seed sets. Value-diversified and boundary strategies yield 289 and 309. All 682 B6 detections arise in its eight planned eligible dtype cases; training/batch diagnostics remain unscored.

\begin{table}[htbp]
\caption{Discovery by corpus and comparison population. Panel (a) shares an oracle; panel (b) retains each external strategy's own decision rule.}
\label{tab:main-results}
\label{tab:external}
\small
\begin{tabular*}{\linewidth}{@{\extracolsep{\fill}}lrrrrr@{}}
\toprule
\multicolumn{6}{@{}l}{\textbf{(a) Shared-protocol findings: count (percentage)}} \\
Strategy & C2 (85) & C3 (227) & C5 (526) & C3+C5 (753) & All (838) \\
\midrule
B1u & 17 (20.0) & 95 (41.9) & 65 (12.4) & 160 (21.2) & 177 (21.1) \\
B3 & 17 (20.0) & 96 (42.3) & 66 (12.5) & 162 (21.5) & 179 (21.4) \\
B4 & 24 (28.2) & 115 (50.7) & 150 (28.5) & 265 (35.2) & 289 (34.5) \\
B5 & 26 (30.6) & 119 (52.4) & 164 (31.2) & 283 (37.6) & 309 (36.9) \\
B6 & 20 (23.5) & 170 (74.9) & 492 (93.5) & 662 (87.9) & 682 (81.4) \\
\Mfull & 30 (35.3) & 170 (74.9) & 492 (93.5) & 662 (87.9) & 692 (82.6) \\
\Mdir & 30 (35.3) & 170 (74.9) & 492 (93.5) & 662 (87.9) & 692 (82.6) \\
\bottomrule
\end{tabular*}
\par\medskip
\begin{tabular*}{\linewidth}{@{\extracolsep{\fill}}lrrrrrr@{}}
\toprule
\multicolumn{7}{@{}l}{\textbf{(b) External findings: counts under the respective oracles}} \\
Strategy & Decidable /838 & C2 & C3 & C5 & C3+C5 & All \\
\midrule
B7 & 628 & 18 & 56 & 115 & 171 & 189 \\
B8 & 643 & 21 & 57 & 144 & 201 & 222 \\
B9 & 496 & 25 & 141 & 308 & 449 & 474 \\
B10 & 778 & 0 & 9 & 41 & 50 & 50 \\
\bottomrule
\end{tabular*}
\end{table}

\Mfull and \Mdir detect the same 692 candidates: all 179 IID-32 detections plus 513 others (3.87 times its yield). On the collected 753-candidate C3+C5 panel, their yield gain over IID-32 is 66.4 percentage points, with a paired task-cluster 95\% interval of [60.8,71.5] over 238 task clusters. Gains over B4 and B5 are 52.7 [46.6,58.7] and 50.3 [44.1,56.4] points. These are panel-specific yield differences, not defect recall.

The 692 discoveries have disjoint stage categories: both strategies flag 58 candidates during preparation (three import, 55 model setup); \Mfull additionally records 409 candidate-invocation failures and 225 observable-behavior mismatches, versus 424 and 210 for \Mdir. The remaining 96 candidates are all-inconclusive because state binding is unusable; 50 have passing observations without a failure.

\paragraph{Observed domains and overlapping discoveries.}
The strategies cover the same 498 accepted candidates but provide different evidence (\cref{tab:witness-domains}): 16 have native witnesses in \Mfull and only dtype witnesses in \Mdir. Native evidence takes precedence within each strategy. Construction and stopping change observation opportunities, so these partitions do not measure latent domain recall.

\begin{table}[htbp]
\caption{Observed witness domains by strategy. Native and dtype-only are disjoint within each strategy; early stopping limits which domains are observed.}
\label{tab:witness-domains}
\small
\begin{tabular*}{\linewidth}{@{\extracolsep{\fill}}lrrrr@{}}
\toprule
 & \multicolumn{2}{c}{All candidates (838)} & \multicolumn{2}{c}{B1u-accepted (548)} \\
\cmidrule(lr){2-3}\cmidrule(lr){4-5}
Strategy & Native & Dtype-only & Native & Dtype-only \\
\midrule
B1u & 177 & 0 & 0 & 0 \\
B3 & 179 & 0 & 3 & 0 \\
B4 & 289 & 0 & 113 & 0 \\
B5 & 309 & 0 & 132 & 0 \\
B6 & 0 & 682 & 0 & 491 \\
\Mfull & 313 & 379 & 136 & 362 \\
\Mdir & 297 & 395 & 120 & 378 \\
\bottomrule
\end{tabular*}
\end{table}

Independently executed B6 and B5 satisfy $D_{B6}\cup D_{B5}=D_{\mathrm{M\text{-}dir}}=D_{\mathrm{M\text{-}full}}$ over all 838 candidates. All ten additions beyond B6 are native C2 discrepancies; B5 finds all ten and B4 nine. This observed union supplies no measured budget or cost for a combined strategy and does not isolate a unique discovery benefit of source fingerprints or fault information.

\paragraph{External strategies.}
\Cref{tab:main-results}(b) reports findings and decidable coverage under each external strategy's inputs, oracle, and validity conditions. Nominal 32-call charges do not equalize decidable comparisons. Sanitizer findings require positive dynamic error/hazard counts. All 838 targets remain in the population, but these results cannot form a common semantic-recall ranking.

\paragraph{Testing cost.}
Charged cost aggregates retained in-scope time rather than total physical execution. It includes original-phase selected observations and each completion stage's in-scope prefix through its first \fail in plan order. Across 838 candidates, \Mfull and \Mdir charge 65.14 and 48.14 GPU-node hours (26.1\% lower; \cref{tab:cost}). Their first-witness medians over the shared 692 detected candidates are 108.67 and 91.45 seconds; at 80 seconds, \Mfull finds 257 versus 234 for \Mdir. Different plans and stopping preclude a controlled reordering interpretation.

Counting all selected in-scope time or all selected time, including diagnostics, gives differences of 25.36\% and 25.41\%, respectively. These scope checks exclude unretained completion attempts and do not recover total physical execution; Appendix~\ref{sec:app-cost} documents record selection and accounting.

\begin{table}[htbp]
\caption{Recorded timing and charged cost. Medians condition on detection; Total is node wall-hours per 1,000 targets, except B10.}
\label{tab:cost}
\small
\begin{tabular*}{\linewidth}{@{\extracolsep{\fill}}llrrr@{}}
\toprule
Strategy & Node & \makecell[r]{Median\\first (s)} & \makecell[r]{Detected\\by 80 s} & Total \\
\midrule
B1u & GPU & 52.1 & 154 & 24.7 \\
B3 & GPU & 52.7 & 150 & 87.0 \\
B4 & GPU & 54.3 & 218 & 112.0 \\
B5 & GPU & 55.7 & 199 & 114.4 \\
B6 & GPU & 52.9 & 638 & 23.7 \\
B7 & GPU & --- & --- & 37.6 \\
B8 & GPU & --- & --- & $\geq121.9$ \\
B9 & CPU & --- & --- & 185.4 \\
B10 & GPU & --- & --- & $89.1^{\dagger}$ \\
\Mfull & GPU & 108.7 & 257 & 77.7 \\
\Mdir & GPU & 91.4 & 234 & 57.5 \\
\bottomrule
\end{tabular*}
\par\smallskip
\raggedright
Dashes denote unavailable comparable timings. B8 has timing for 834/838 targets, hence a lower bound. $\dagger$B10 uses 69.32 recorded hours over 778 executed candidates (89.1 h/1k), or 82.7 h/1k over 838 targets without imputing the other 60. Other rows use 838. Appendix~\ref{sec:app-cost} gives first-call and remaining costs; these do not isolate initialization and execution.
\end{table}

\Needspace{7\baselineskip}
B6 finds 682 candidates at 23.7 h/1k, including 638 by 80 seconds. Its recorded dtype prefix retains all 682 at 17.9 h/1k over 838 targets; standalone prefix execution was not measured. B6's total includes diagnostics excluded by \Mfull/\Mdir accounting. B7/B9 timings cover all 838 attempts, including failures; whole-batch durations cannot recover their first-failure timestamps.

\paragraph{Input schedule versus oracle.}
The oracle comparison uses 668 candidates selected by a prior raw \texttt{kill} flag in the anchor or \Mfull, regardless of final score eligibility (170 C3, 498 C5; \cref{tab:oracle}). The other 85 candidates in the 753-candidate panel did not meet this condition and were not selected for fixed-tolerance reruns. Within this selected cohort, the dtype-aware anchor detects all 109 fixed-oracle discoveries plus 51 others, and both full-suite executions detect the same 661 candidates. Higher full-suite yield thus persists with fixed tolerance in this cohort; it does not estimate the oracle's contribution to gains over the full population.

\begin{table}[htbp]
\caption{Oracle sensitivity on 668 candidates selected by prior raw \texttt{kill} flags, using aligned plans from independent executions. Dtype-aware tolerances tighten FP32 to $10^{-4}$; FP16/BF16 remain at $10^{-2}$.}
\label{tab:oracle}
\small
\begin{tabularx}{\linewidth}{@{}Yrr@{}}
\toprule
Input schedule & Fixed $10^{-2}$ & Dtype-aware \\
\midrule
Five-draw IID & B1: 109/668 & B1u: 160/668 \\
Full suite, up to 32 calls & M-fixed: 661/668 & \Mfull: 661/668 \\
\bottomrule
\end{tabularx}
\end{table}

Case-level verdicts differ despite equal full-suite discovery sets (Appendix~\ref{sec:app-oracle}). These are matched cases from independent executions, not rescored saved outputs; stopping also changes which cases are observed.
\Needspace{6\baselineskip}
\rqanswer{Directed construction retains the full suite's discovery set at lower charged in-scope cost, with different witness domains. The baseline overlap precludes claiming a unique discovery benefit of fault information. Broader testing retains its yield gain under both tolerances on the selected oracle cohort.}

\subsection{RQ3: Independent Labels and Control Alarms}
\label{sec:rq3}
The C6 campaign adapts eleven strategies to gpuemu's ten seeded bugs and 16 controls. FP64 reference outputs are rounded to the input dtype. Ordinary adaptations use the operator/dtype metadata as both absolute and relative tolerances, with the relative term scaled by the reference; B1 fixes both at $10^{-2}$, whereas B9 uses the metadata as an absolute bound only. Thus the adaptations do not share one numerical gate. B6 uses a dtype/shape grid. Any failure flags a subject; every subject--strategy pair has a conclusive observation.

All eleven strategies detect all ten bugs (\cref{tab:controls}), so these saturated results cannot establish superior recall. \Mdir flags two of 16 controls versus seven for \Mfull; several baselines raise none. Both plan 32 calls per subject. \Mfull records 806 decidable observations; \Mdir records 728, with 52 additional BF16 cases inconclusive because input conversion fails before kernel invocation.

\begin{table}[htbp]
\caption{Seeded-bug detection and control alarms in the adapted C6 campaign. Each row tests ten seeded bugs and 16 controls; control labels remain unchanged.}
\label{tab:controls}
\small
\begin{tabularx}{\linewidth}{@{}Yrr@{}}
\toprule
Strategy & Bugs detected /10 & Controls flagged /16 \\
\midrule
B1, B1u, B3, B4, B6, B7, B9 (each) & 10 & 0 \\
B5 Boundary values & 10 & 7 \\
B8 KernelBenchX-style & 10 & 1 \\
\Mfull & 10 & 7 \\
\Mdir & 10 & 2 \\
\bottomrule
\end{tabularx}
\par\smallskip\raggedright
Anchors use their gpuemu adaptations. Compute Sanitizer and M-fixed are not additional C6 rows.
\end{table}

\Needspace{5\baselineskip}
Control alarms have unresolved numerical significance. \Mfull's seven flagged controls yield 14 FP32 near-overflow failures: six NaN-pattern mismatches and eight tolerance violations. \Mdir's three FP32 alarms involve all-negative/reduction-adversarial \nolinkurl{attention_triton} and reduction-adversarial NumPy \nolinkurl{matmul}. Absent outputs and elementwise errors prevent further adjudication; independent labels remain unchanged. Appendix~\ref{sec:app-controls} gives policies, tolerances, and reconstructed shapes.
\rqanswer{\Mdir flags fewer controls than \Mfull with the same saturated bug detection. Unequal test exposure prevents isolating a guidance effect; the results establish neither zero false positives nor superiority to zero-alarm baselines.}

\subsection{RQ4: Controlled Faults and Test-Condition Transfer}
\label{sec:rq4}
\Cref{tab:c1} separates excluded probes, harness-scoped equivalents, anchor kills, witnessed survivors, and unresolved cases. Appendix Sections~\ref{sec:app-mutations}--\ref{sec:app-witnesses} give operator, control, and witness-condition details.

\begin{table}[htbp]
\caption{Complete C1 dispositions. The 77 witnessed survivors comprise 68 screening, seven enhanced-search, and two additional FP16 witnesses.}
\label{tab:c1}
\label{tab:c1-stages}
\label{tab:c1-mechanisms}
\small
\begin{tabularx}{\linewidth}{@{}lrY@{}}
\toprule
Disposition & Probes & Evidence and scope \\
\midrule
Excluded & 571 & 105 compile/stillborn; 441 host-control failures; 25 other unusable comparisons \\
Anchor-killed & 539 & Mutant--reference failures under the anchor \\
Harness-equivalent & 9 & Static dependency evidence; anchor not executed \\
Witnessed survivors & 77 & Host-relative differences in the stated search domains \\
Unresolved & 450 & No established equivalence or distinguishing witness \\
\midrule
Total & 1,646 & All generated probes \\
\bottomrule
\end{tabularx}
\end{table}

Of 459 probes without a screening witness, 453 have enhanced-search records; six lack an executable plan. After the two additional FP16 witnesses, 450 remain unresolved. Searches compare mutant with host (\cref{sec:study}), so a witness need not be a same-reference anchor failure. Timeouts and incomplete plans do not establish equivalence.

There are 616 probes with established non-equivalence: 539 anchor kills and 77 witnessed survivors. The conditional fraction not exposed by the anchor is $77/616=12.5\%$ over the expanded witness domain, including two FP16 cases beyond the FP32 anchor. It is not a same-domain FP32 false-negative rate. Excluding 450 unresolved probes restricts that estimate rather than resolving them.

Saved witness and trial fields partition the 77 probes into 69 structured-value, five random-input, two additional FP16 cases, and one confirmed crash without a template identity. These observation categories do not establish mutually exclusive causes: random-input witnesses do not prove tolerance absorption, and comparison-pair changes limit input or precision attribution.

\paragraph{Dependence on prior support.}
The last-element template occurs twice in all 838 natural-candidate \Mdir plans, with 716 selected scored observations across 565 candidates: 503 \pass, 211 \inconclusive, and two \fail. Masking its observations retains 690/692 discoveries and 295/297 native discoveries. The two affected candidates each have a one-element FP32 tolerance violation (Appendix~\ref{sec:app-recommendations}). This masking analysis measures fixed-witness dependence on the template, not deletion of one association: other classes and general fill can select it, and association-level construction traces are unavailable.

\paragraph{Task-held-out recommendations.}
Under \cref{sec:study}'s confirmed-evidence protocol, class-conditioned recommendations cover $48/75=64.0\%$ of the primary witnesses. Seven probes have insufficient coverage evidence; 20 have conclusive records for every recommended template without a witness among them. The missing-template probe remains in the denominator. Global top-eight covers 51, with insufficient evidence for the remaining 24.

\begin{table}[htbp]
\caption{Offline recommendation evidence on 75 held-out probes. ``No witness'' means all recommended templates have conclusive records without a witness; unknown means insufficient evidence. Recommendation totals are not executed calls.}
\label{tab:crossfit}
\small
\begin{tabular*}{\linewidth}{@{\extracolsep{\fill}}lrrrrrrrr@{}}
\toprule
Per-probe cap & 1 & 2 & 3 & 4 & 5 & 6 & 7 & 8 \\
\midrule
Class covered /75 & 11 & 29 & 46 & 46 & 48 & 48 & 48 & 48 \\
Class no witness & 20 & 20 & 20 & 20 & 20 & 20 & 20 & 20 \\
Class unknown & 44 & 26 & 9 & 9 & 7 & 7 & 7 & 7 \\
Global covered /75 & 0 & 3 & 22 & 35 & 35 & 41 & 48 & 51 \\
Global no witness & 6 & 5 & 2 & 2 & 2 & 1 & 0 & 0 \\
Global unknown & 69 & 67 & 51 & 38 & 38 & 33 & 27 & 24 \\
Class recommendations & 75 & 146 & 192 & 216 & 228 & 228 & 228 & 228 \\
Global recommendations & 75 & 150 & 225 & 300 & 375 & 450 & 525 & 600 \\
\bottomrule
\end{tabular*}
\end{table}

Observation availability differs across the rankings. At cap three, class lists have conclusive records for 139/192 recommendations versus 104/225 globally. To assess ordering on shared evidence, we intersect each probe's two top-eight lists with its conclusive templates, retain both rankings, and take the first $k$. Among 73 probes with nonempty pools, class/global coverage is 28/8 at $k=1$, 45/47 at $k=2$, and 47/47 at $k=3$. Pools contain at most four templates, making equality at larger $k$ inevitable. This conditioning excludes unique and unobserved templates and does not correct informative missingness or early stopping. Only six probes have fully conclusive records for both unchanged cap-one lists, and none at caps seven or eight (Appendix~\ref{sec:app-recommendations}).
\rqanswer{Witness conditions and recorded coverage are measurable, but the ranking comparison depends on observation support. Task-held-out matching and its conditional sensitivity do not establish deployed detection or runtime gains.}

\subsection{RQ5: Component Contribution and Dtype}
\label{sec:rq5}
The removal study conditions on 609 previously detected C3+C5 candidates: 165 C3 and 444 C5, comprising 401 level-1 and 208 level-2 implementations. Each variant removes one dimension and applies its remaining-suite and budget-fill construction. \Cref{tab:ablation} measures retention on this selected cohort, not full-population discovery.

\begin{table}[htbp]
\caption{Component analysis on 609 detected candidates. The first three numeric columns describe executed removal-and-refill constructions. The last masks observed \Mfull witnesses, without constructing or executing new tests.}
\label{tab:ablation}
\small
\begin{tabularx}{\linewidth}{@{}Yrrrr@{}}
\toprule
Construction & Detections & \makecell[r]{Retention\\(\%)} & \makecell[r]{Incon-\\clusive} & \makecell[r]{Candidates after\\masking} \\
\midrule
\Mfull & 609 & 100.0 & 0 & 609 \\
Without value & 609 & 100.0 & 0 & 339 \\
Without dtype & 269 & \makecell[r]{44.2\\(at most 44.8)} & 4 & 270 \\
Without training & 607 & 99.7 & 0 & 609 \\
Without repeat & 606 & 99.5 & 0 & 609 \\
Without configuration & 602 & 98.9 & 0 & 609 \\
\bottomrule
\end{tabularx}
\end{table}

Witness masking asks which observed failures depend on a dimension. The fixed \Mfull evidence contains 270 value-domain and 339 dtype-domain candidates; masking either retains the other. No other category contributes scored witnesses. Early stopping leaves tests unobserved, so masking cannot predict a reconstructed plan.

Executed removal-and-refill changes testing opportunities. Removing dtype loses 340 detections; even if all four inconclusive cases yielded witnesses, at least $336/609=55.2\%$ remain lost. Losses after removing training, repeat, or configuration do not isolate those dimensions: their diagnostics are unscored, while reconstruction changes eligible tests.

\begin{table}[htbp]
\caption{Dtype cases with native controls and execution evidence. Additional-dtype support is distinguished from the publisher's original guarantee.}
\label{tab:dtype-cases}
\small
\begin{tabularx}{\linewidth}{@{}p{25mm}YY@{}}
\toprule
Candidate / task & FP32 control & Additional-dtype observation \\
\midrule
C3-23530 / LeNet-5 & 48 applicable inputs agree; four custom ReLU launches & FP16/BF16 reject the first input at a Float pointer check, before custom launch \\
C3-00995 / Matmul & 32 original-distribution inputs agree & FP16 CUDA executes; 723,115 boundary elements exceed tolerance (max absolute difference 1.0), masked by readout; BF16 rejected \\
\bottomrule
\end{tabularx}
\end{table}

C3-23530's dispatch rejection in \cref{tab:dtype-cases} also occurs in the main campaign after 21 passing FP32 policies, with a valid reference. For C3-00995, scalar-type accumulation is consistent with precision loss, but a unique repair-based cause is unestablished; its FP32 large-value witness precludes a dtype-exclusive attribution. Appendix~\ref{sec:app-dtype} gives additional case evidence.
\rqanswer{Dtype is necessary to retain this suite's observed discoveries on the selected population. Passing FP32 controls neither exercise low-precision dispatch nor establish accurate computation after that dispatch succeeds.}

\subsection{Replacement Study: Binding and Execution Boundaries}
\label{sec:replacement}
\paragraph{Design and controls.}
We attempt 838 candidates in three requested dtypes, totaling 2,514 configurations. Network candidates retain their model; applicable operators use a deterministic shallow readout, such as ReLU and adaptive average pooling to eight features. Unsuitable cases retain explicitly identified direct comparisons. Inputs are synthetic tensors, not labeled images; outcomes concern tensors, not classification accuracy. Each configuration keeps its shape and plans at most 32 original-distribution plus 16 applicable structured inputs, with early stopping and a 300-second timeout.

Candidate and reference receive synchronized corresponding state and identical inputs. C3 uses the paired functional model's default Python path as reference and compiled entry as candidate; C2/C5 use a reference copy and synchronized published implementation. Execution uses evaluation mode, no gradients/autocast, and disabled TF32. Absolute/relative tolerances are $(10^{-2},10^{-2})$ for FP32/FP16 and $(10^{-2},5\times10^{-2})$ for BF16, separate from the main oracle.

Native-branch guards, explicit attachment of an existing custom entry, or direct execution evidence address fallback; published algorithms are not repaired. Requested and actual dtype, custom execution, and comparison validity are separate observations. Reference/state controls support the specified bindings without proving every adapter correct; attribution also uses source, execution traces, and reproducible witnesses.

\begin{table}[htbp]
\caption{All 2,514 replacement outcomes by corpus and requested dtype. Agreement requires evidence of custom execution. Every terminal category remains in the denominator.}
\label{tab:replacement}
\small
\begin{tabular*}{\linewidth}{@{\extracolsep{\fill}}llrrrrrrr@{}}
\toprule
Corpus & Dtype & Configs & Agree & \makecell[r]{Final\\diff.} & Runtime & \makecell[r]{Local/\\direct} & \makecell[r]{Dtype\\reject} & Other \\
\midrule
C2 & FP32 & 85 & 4 & 3 & 0 & 1 & 0 & 77 \\
C2 & FP16 & 85 & 0 & 2 & 0 & 0 & 6 & 77 \\
C2 & BF16 & 85 & 0 & 1 & 0 & 0 & 7 & 77 \\
C3 & FP32 & 227 & 135 & 9 & 1 & 5 & 0 & 77 \\
C3 & FP16 & 227 & 7 & 0 & 0 & 1 & 138 & 81 \\
C3 & BF16 & 227 & 0 & 0 & 0 & 0 & 148 & 79 \\
C5 & FP32 & 526 & 496 & 3 & 1 & 7 & 0 & 19 \\
C5 & FP16 & 526 & 0 & 0 & 0 & 0 & 508 & 18 \\
C5 & BF16 & 526 & 0 & 0 & 0 & 0 & 508 & 18 \\
\midrule
Total & --- & 2,514 & 642 & 18 & 2 & 14 & 1,315 & 523 \\
\bottomrule
\end{tabular*}
\end{table}

\paragraph{Complete outcomes.}
The 18 final-output discrepancies and two runtime failures involve 17 candidates. Fourteen local/direct differences comprise nine masked downstream and five without a readout. Appendix Table~\ref{tab:app-other-outcomes} enumerates the 19 subcategories of the 523 other outcomes. Agreement remains finite evidence.

\paragraph{Evidence beyond main-study non-detection.}
Among 146 candidates without a scored \Mfull failure, 96 are all-inconclusive and 50 have only passing observations of the tested callable. Joining stable identities yields the four FP32 final-output/runtime witnesses in \cref{tab:replacement-cases}. The first three require binding changes; C3-17320 also changes reference. C5-00631's failure at input 37 therefore does not establish budget as its main-study obstacle. Four further candidates have only downstream-masked differences and remain outside this category.

\begin{table}[htbp]
\caption{All four FP32 final-output/runtime witnesses among candidates without a scored \Mfull failure. Cross-study changes restrict false-negative attribution.}
\label{tab:replacement-cases}
\small
\begin{tabularx}{\linewidth}{@{}p{21mm}YY@{}}
\toprule
Candidate & Replacement witness & Comparison scope \\
\midrule
C3-17320 & Linear--Sigmoid--Sum--LogSumExp returns $[1]$ instead of scalar $[]$ & Main binding inconclusive; recovered reference uses a different operation sequence \\
C3-20564 & GEMM--GroupNorm--Min--Bias readout returns $[128,8]$ instead of $[1,8]$ & Main state binding inconclusive; replacement binding permits comparison \\
C5-00631 & LayerNorm on constant-0.1 input: final max absolute difference $\approx0.128$; first failure at test 37 & Main parameter binding inconclusive \\
C5-01485 & MobileNetV1 custom Conv+BN+ReLU returns three dimensions; next layer raises dimension error & Main \pass runs PyTorch-only \texttt{forward}; replacement executes the dormant custom entry \\
\bottomrule
\end{tabularx}
\end{table}

\paragraph{A dormant custom implementation.}
C5-01485's 24 main \pass observations exercise PyTorch-only \texttt{forward}. Its custom function returns $[32,224,224]$ instead of $[10,32,112,112]$: the wrapper allocates \texttt{\{out\_channels, height, width\}}, omitting batch and stride reduction. One custom launch and passing reference/state controls support this attribution (Appendix~\ref{sec:app-replacement}).

Numerical strictness cannot repair bindings or invoke an unattached function. Replacement witnesses remain separate from main counts: a replacement claim needs the intended custom code executing with usable bindings and explicit input obligations.

\section{Threats to Validity}
\label{sec:threats}
\paragraph{Construct validity.}
Study contracts preset input domains, determinism, and tolerances without establishing publisher obligations. We distinguish native FP32, additional-dtype, and external-tool findings. Yield includes preparation and invocation failures alongside comparison mismatches, so it is not defect recall. C6 control alarms are unadjudicated and do not establish precision on natural candidates (\cref{sec:rq3}).

\paragraph{Internal validity.}
Protocol choices affect attribution even with reference/state controls. Replacement witnesses have distinct bindings and execution paths (\cref{sec:replacement}). The unsupported prior association is localized, but template masking cannot identify the effect of removing that association on plan construction (\cref{sec:rq4}). No executed source-agnostic strategy holds eligible budget, dtype allocation, and stopping constant; guidance and ordering effects remain unisolated. C6 exposure differs, and removal-and-refill conditions on 609 detected candidates.

\paragraph{External validity.}
The natural collections are nonrandom KernelBench-derived samples with incomplete generation provenance and task overlap with C1. Task-held-out cross-fitting measures confirmed-probe template coverage, not task-disjoint natural-candidate performance. Host selection, exclusions, and unresolved probes restrict witness-condition frequencies; C6 is a small adapted seeded corpus. Fixed shapes, synthetic inputs, and selected replacement networks leave broader workloads and GPU architectures untested. Replacement results establish neither online optimization convergence nor end-to-end search-cost benefits.

\paragraph{Conclusion validity.}
Task-cluster intervals address dependence, not execution variability or nonrandom selection. Prior raw \texttt{kill} selection and independent executions limit oracle-panel inferences. Offline rankings aggregate seeds over partial observations; conditioning on shared observable templates does not remove selection from stopping or timeouts. Probe classes are known, whereas natural fingerprints supply hypotheses. Source-ledger checks bound original-record selection, but unretained completion attempts and global main-study precision settings remain unverified. Charged costs describe retained observations, and first-witness medians condition on detection. Confirmations add no independent units.

\section{Related Work}
\label{sec:related}
\paragraph{Kernel generation and validation.}
KernelBench~\cite{ouyang2025kernelbench} compares sampled outputs with PyTorch and measures runtime; released CUDA collections~\cite{cudaL1,aicudaengineer,kernelbenchsamples} supply our candidates. robust-kbench~\cite{robustkbench} varies initialization and configuration and uses LLM screening. KernelBenchX~\cite{kernelbenchx} separates call validity, output agreement, and efficiency; KernelBench-Verified~\cite{kernelbenchverified} adds hidden distributions and a TF32-enabled baseline. Our strategy ports retain their respective inputs and oracles.
Li et al.~\cite{li2026correctbutslow} study Triton and TileLang kernels that pass numerical checks yet run substantially slower than library baselines, separating replacement efficiency from numerical agreement. Our evaluation concerns validation outcomes and testing cost, not the execution efficiency of replacement kernels.

\paragraph{GPU and tile-program bug characterization.}
Rathnasuriya et al.~\cite{rathnasuriya2025gpunumerical} characterize 397 GPU numerical bugs and develop GPU-NBDetect using bug-triggering inputs and numerical oracles. Their tile-program study~\cite{rathnasuriya2026tilebugs} analyzes 301 code-generation bugs, including dependencies on tensor shapes, types, and backend transformations. These studies characterize observed defects; our controlled probes measure which faults survive a specified validator and which test conditions expose them.

\paragraph{Oracle measurement and candidate audit.}
Du et al.~\cite{measuringchecker} combine mutation analysis, validity checks, per-problem suite selection, and taxonomy-informed generation, holding out mutants within each problem. Our analysis holds out whole tasks and compares class-conditioned with global rankings under partial observation. Shah and Shrestha~\cite{contractgrade} audit accepted kernels using applicable contract gates and positive controls. Our findings connect equal candidate yield with different witness domains, recommendation limits with observation coverage, and replacement verdicts with binding and custom execution.

\paragraph{Numerical testing and verification.}
The gpuemu studies examine high-precision references~\cite{gpuemu}, tolerance calibration~\cite{gpuemutolerance}, and input generation~\cite{gpuemuinputs}; we retain their labels when assessing alarms. ProofWright~\cite{proofwright} proves memory safety and race freedom under inferred preconditions, with semantic equivalence for restricted element-wise kernels. Compute Sanitizer~\cite{nvidiasanitizer} detects dynamic safety hazards. These guarantees differ from the numerical witnesses measured here.

\paragraph{Mutation analysis and fault-informed testing.}
MUTGPU~\cite{mutgpu} targets CUDA C; CLTestCheck~\cite{cltestcheck} assesses OpenCL tests through coverage, fault seeding, and schedule amplification. Mu2~\cite{vikram2023guiding} uses online mutation feedback during Java greybox fuzzing. Our source-guided selection uses an offline prior without mutating new candidates. Equivalent-mutant techniques include compiler checks~\cite{papadakis2015tce} and targeted suppression~\cite{kushigian2024equivalent}; we retain the distinction between harness-scoped equivalence and unresolved survival~\cite{madeyski2013overcoming}.

\paragraph{Uncertainty in code models.}
CodeImprove~\cite{rathnasuriya2025codeimprove} combines a model-derived input-validity score with semantics-preserving program transformations to improve code-model predictions. A study of code generation finds that confidence and entropy provide incomplete, model-dependent evidence of execution failure~\cite{rathnasuriya2026overconfidence}. For defect prediction and vulnerability detection, improved calibration scores do not consistently improve selective prediction~\cite{rathnasuriya2026beyondconfidence}. These approaches assess or improve model predictions; \ourtool records execution-based witnesses under explicit input and oracle contracts. Our \inconclusive verdict denotes missing valid execution evidence, rather than a threshold on model confidence.

\Needspace{6\baselineskip}
\section{Conclusion}
\label{sec:conclusion}
\ourtool shows that sampled acceptance leaves native discrepancies and additional-dtype limitations unchecked, while wrapper agreement can leave custom code untested. Recommendation comparisons depend on both ranking and observable evidence; shared-template analyses limit generalization from recorded coverage. Replacement decisions therefore need witness domains, comparison contracts, and actual execution paths alongside candidate yield and charged cost.

\begin{acks}
This work was partially supported by Fundamental and Interdisciplinary Disciplines Breakthrough Plan of the Ministry of Education of China (No. JYB2025XDXM118) and National Natural Science Foundation of China under Grant Nos. U25A6023 and 92464301.

Tao Xie is also affiliated with Key Laboratory of High Confidence Software Technologies (Peking University), Ministry of Education, China; Shanghai Institute of Systems for Open Computing, China.
\end{acks}

\bibliographystyle{ACM-Reference-Format}
\bibliography{ref}
\clearpage
\appendix
\section{Supplementary Evidence}
\label{sec:appendix-evidence}
\setcounter{table}{0}
\renewcommand{\thetable}{S\arabic{table}}
\renewcommand{\theHtable}{supp.\arabic{table}}
\setlength{\heavyrulewidth}{1pt}
\setlength{\lightrulewidth}{0.65pt}
\setlength{\cmidrulewidth}{0.5pt}
\setlength{\LTcapwidth}{\textwidth}

This appendix provides witness configurations, record-selection details, and additional breakdowns supporting the main results.
The study populations, validation contracts, and aggregate comparisons are specified in \cref{sec:background,sec:study,sec:method,sec:evaluation}.

\Needspace{10\baselineskip}
\subsection{Native-Witness Details}
\label{sec:app-native}

The following configurations and recorded symptoms supplement the native-witness analysis in \cref{sec:rq1}.
The accepted candidates with native FP32 witnesses comprise 100 level-1, 28 level-2, and eight level-3 implementations.
Across the controlled strategies' observed native tests, 95 candidates have only finite-value tolerance violations, 30 only NaN/Inf-position mismatches, and 11 both.
These are symptom categories, not source-level causes.
The three cases below use FP32, evaluation mode, and seed zero; their original observations are bound to candidate/reference source hashes.

For \nolinkurl{C2__KB_L1_P19__d0640b9cce2e0f4a}, the large-magnitude input has shape $[4096,393216]$, with gradients disabled.
The post-call input comparison reports 805,319,693 of 1,610,612,736 elements above the threshold and maximum absolute difference 5767.48.

For \nolinkurl{C2__KB_L1_P37__f84d95d3af89255b}, denormal inputs of shape $[112,64,512,512]$ produce Inf-position differences at all 1,879,048,192 output elements.
This size exceeds the candidate's 1,048,576-element custom-CUDA guard.
The saved message neither identifies which side contains Inf nor establishes the reference's numerical validity in this domain.

For \nolinkurl{C2__KB_L2_P93__683a5034e986671e}, a large-magnitude input yields 403 threshold violations among 276,889,600 output elements and a reported maximum absolute difference of $2.28085\times10^{-4}$.
Its cause remains unadjudicated.
The finite comparisons use $\mathrm{atol}=\mathrm{rtol}=10^{-4}$.
Counts and rounded error magnitudes come from saved oracle messages; no saved-output rescoring or numerical replay underlies these descriptions.

The three C2 additional-dtype cases that return discrepant outputs consist of one output-dtype mismatch, one finite-value tolerance violation, and one NaN-position mismatch.
Failure-stage classification uses recorded stages and oracle verdicts; comparison failures are retained when an ancillary state-restore error also occurs.

\Needspace{10\baselineskip}
\subsection{Cost Components and Record Reconstruction}
\label{sec:app-cost}

Table~\ref{tab:app-cost-components} gives the recorded cost components underlying the totals in \cref{tab:cost}.
First-call cost sums the full wall time of each candidate's first invocation in the original ledger; the remaining cost is charged Total minus First call.
These components do not isolate initialization and execution.
Only strategies with available component measurements are listed.

\Needspace{20\baselineskip}
\begingroup
\small
\setlength{\tabcolsep}{4pt}
\begin{xltabular}{\linewidth}{@{}Xrr@{}}
\caption{First-call and remaining charged costs, in GPU-node wall-hours per 1,000 targets.}\label{tab:app-cost-components}\\
\toprule
\textbf{Configuration} & \makecell[r]{\textbf{First-call}\\\textbf{cost}} & \makecell[r]{\textbf{Remaining}\\\textbf{charged cost}} \\
\midrule
\endfirsthead
\caption[]{First-call and remaining charged costs, in GPU-node wall-hours per 1,000 targets. (continued)}\\
\toprule
\textbf{Configuration} & \makecell[r]{\textbf{First-call}\\\textbf{cost}} & \makecell[r]{\textbf{Remaining}\\\textbf{charged cost}} \\
\midrule
\endhead
\midrule
\multicolumn{3}{r@{}}{\textit{Continued on next page}}\\
\endfoot
\bottomrule
\endlastfoot
B1u anchor & 13.4 & 11.3 \\[4pt]
B3 IID-32 & 13.4 & 73.7 \\[4pt]
B4 Value-diversified & 14.8 & 97.2 \\[4pt]
B5 Boundary values & 14.9 & 99.5 \\[4pt]
B6 Composite execution-context arm & 13.6 & 10.0 \\[4pt]
M-full & 14.8 & 62.9 \\[4pt]
M-dir & 17.3 & 40.2 \\[4pt]
\end{xltabular}
\endgroup
\par\medskip

\paragraph{Selection and ledger fields.}
Original records are deduplicated by candidate and test identity, preferring \fail.
Within each completion stage, selected in-scope observations follow original plan order through its first \fail.
Out-of-scope calls, superseded duplicate attempts, and completion calls beyond that prefix are excluded; charged failures, timeouts, and inconclusive observations remain.
The ledger binds each selected observation to candidate, strategy, case identity, phase, source order, wall time, charge flag, and exclusion reason.

Six provenance-matched source ledgers contain 88,547 records with 88,547 distinct candidate--strategy--case keys.
First-recorded and \fail-preferring selection therefore choose identical original records; this agreement is checked on source ledgers, not inferred from the deduplicated export.
Candidate and contract hashes match registered subjects, and directed records match the saved plans.
File order is available, but global physical timestamps are not.

\paragraph{Skipped-slot completion.}
Completion supplies 214 previously unobserved cases and fills 37 timeout-stop slots.
Each replaced slot has zero wall time, zero observed candidate invocations, and the reason ``skipped after 5 consecutive parent-timeout cases in this subject/strategy pair.''
These are unexecuted placeholders, not competing earlier executions.
The 18 \Mfull and 19 \Mdir slots belong to one candidate; paired case and contract identities agree.
Retained completion gives 16 inconclusive and two \fail outcomes for \Mfull, and 15 inconclusive and four \fail outcomes for \Mdir.
Unselected completion ledgers are outside this audit, so it does not establish the absence of all unrecorded repeated attempts.

\Needspace{8\baselineskip}
\begingroup
\small
\setlength{\tabcolsep}{4pt}
\begin{xltabular}{\linewidth}{@{}Z{1.05}R{1.0}R{0.9}R{0.85}R{0.95}R{1.25}@{}}
\caption{Original-record and completion-slot accounting.}\label{tab:app-record-selection}\\
\toprule
\textbf{Strategy} & \textbf{Original records} & \textbf{Skipped slots} & \textbf{New cases} & \textbf{Filled slots} & \textbf{Selected records} \\
\midrule
\endfirsthead
\caption[]{Original-record and completion-slot accounting. (continued)}\\
\toprule
\textbf{Strategy} & \textbf{Original records} & \textbf{Skipped slots} & \textbf{New cases} & \textbf{Filled slots} & \textbf{Selected records} \\
\midrule
\endhead
\midrule
\multicolumn{6}{r@{}}{\textit{Continued on next page}}\\
\endfoot
\bottomrule
\endlastfoot
M-full & 13,179 & 18 & 26 & 18 & 13,205 \\
M-dir & 9,294 & 19 & 188 & 19 & 9,482 \\
\end{xltabular}
\endgroup
\par\medskip

\paragraph{Alternative time scopes.}
Table~\ref{tab:app-cost-scopes} uses the same selected observation export throughout.
The second row includes all in-scope completion records without the charged-prefix cutoff; the third also includes diagnostics.
These are alternative time scopes, not selections among unretained attempts or measurements of full physical execution.
The CPU reconstruction does not replay GPU candidates.

\Needspace{8\baselineskip}
\begingroup
\small
\setlength{\tabcolsep}{4pt}
\begin{xltabular}{\linewidth}{@{}Z{1.6}R{0.8}R{0.8}R{0.8}@{}}
\caption{Recorded hours under the alternative accounting scopes.}\label{tab:app-cost-scopes}\\
\toprule
\textbf{Selected-record scope} & \textbf{M-full hours} & \textbf{M-dir hours} & \textbf{Relative difference} \\
\midrule
\endfirsthead
\caption[]{Recorded hours under the alternative accounting scopes. (continued)}\\
\toprule
\textbf{Selected-record scope} & \textbf{M-full hours} & \textbf{M-dir hours} & \textbf{Relative difference} \\
\midrule
\endhead
\midrule
\multicolumn{4}{r@{}}{\textit{Continued on next page}}\\
\endfoot
\bottomrule
\endlastfoot
Charged prefixes & 65.14337 & 48.14322 & 26.10\% \\
All in-scope records & 65.20060 & 48.66410 & 25.36\% \\
All records, including diagnostics & 66.75913 & 49.79841 & 25.41\% \\
\end{xltabular}
\endgroup
\par\medskip

\Needspace{10\baselineskip}
\subsection{Oracle-Control Selection and Matched Cases}
\label{sec:app-oracle}

The selection rule and candidate-level results are reported in \cref{sec:rq2,tab:oracle}.
The unselected records comprise 57 C3 and 28 C5 candidates.
Their preliminary anchor dispositions are 82 state-binding failures (57 C3, 25 C5), one C5 parent-process timeout, and two C5 passes with no preliminary \Mfull kill.
These are selection-time dispositions, not outcomes of unexecuted fixed-tolerance controls.

Matching cases by candidate identity, policy, seed, mode, parameters, and replicate yields 2,706 anchor pairs and 8,802 suite pairs across independent executions.
Anchor matches contain 51 \pass-to-\fail transitions and none in reverse.
Suite matches contain 58 \pass-to-\fail, one \fail-to-\pass, and seven \inconclusive-to-\pass transitions.
Saved outputs were not rescored, and the different stopping points affect which cases can be matched.

\Needspace{10\baselineskip}
\subsection{Control-Alarm Configurations}
\label{sec:app-controls}

For the C6 comparison in \cref{sec:rq3}, the ordinary adaptation calls \texttt{np.allclose(candidate, ideal)}, scaling the relative term by the reference result.
The 286 subject--strategy outcomes aggregate 7,670 observations, with at least one \pass or \fail observation for every pair.

The three \Mdir alarm records use FP32 and shape variant zero.
For \nolinkurl{attention_triton}, $\mathrm{atol}=\mathrm{rtol}=10^{-3}$; for the NumPy \nolinkurl{matmul} case, $\mathrm{atol}=\mathrm{rtol}=5\times10^{-2}$.
All three are tolerance violations against the corpus reference, not NaN-pattern mismatches.
Table~\ref{tab:app-control-configurations} gives their seeds and shapes, reconstructed deterministically from the saved runner and metadata rather than recovered from logged tensors.

\Needspace{8\baselineskip}
\begingroup
\small
\setlength{\tabcolsep}{4pt}
\begin{xltabular}{\linewidth}{@{}Z{0.85}Z{0.9}R{0.6}Z{1.65}@{}}
\caption{Seeds and reconstructed shapes for the three M-dir control alarms.}\label{tab:app-control-configurations}\\
\toprule
\textbf{Control} & \textbf{Input policy} & \textbf{Seed} & \textbf{Reconstructed shapes} \\
\midrule
\endfirsthead
\caption[]{Seeds and reconstructed shapes for the three M-dir control alarms. (continued)}\\
\toprule
\textbf{Control} & \textbf{Input policy} & \textbf{Seed} & \textbf{Reconstructed shapes} \\
\midrule
\endhead
\midrule
\multicolumn{4}{r@{}}{\textit{Continued on next page}}\\
\endfoot
\bottomrule
\endlastfoot
\nolinkurl{attention_triton} & All negative & 2093511732 & $Q:[1,8]$; $K,V:[256,8]$ \\
\nolinkurl{attention_triton} & Reduction adversarial & 505391955 & $Q:[8,8]$; $K,V:[256,8]$ \\
\nolinkurl{matmul} & Reduction adversarial & 1239010171 & $A:[2,256]$; $B:[256,64]$ \\
\end{xltabular}
\endgroup
\par\medskip

The original rows retain categorical failures but no output tensors, maximum errors, or affected-element counts.
These quantities remain unknown.
Available corpus-source hashes identify recovered files, not the source bytes at historical execution time.
The configurations identify reported alarms without adjudicating their numerical significance.

\Needspace{10\baselineskip}
\subsection{Controlled-Mutation Details}
\label{sec:app-mutations}

Table~\ref{tab:app-fault-families} expands the aggregate dispositions in \cref{tab:c1} by fault family.

\Needspace{8\baselineskip}
\begingroup
\small
\setlength{\tabcolsep}{4pt}
\begin{xltabular}{\linewidth}{@{}Z{1.6}R{0.7}R{0.9}R{1.0}R{0.8}@{}}
\caption{C1 probe dispositions by fault family; aggregate totals are in Table~\ref{tab:c1}.}\label{tab:app-fault-families}\\
\toprule
\textbf{Disposition} & \textbf{A: arithmetic} & \textbf{B: GPU parallelism} & \textbf{C: ML numerical behavior} & \textbf{D: Broadcast/\allowbreak{}layout} \\
\midrule
\endfirsthead
\caption[]{C1 probe dispositions by fault family; aggregate totals are in Table~\ref{tab:c1}. (continued)}\\
\toprule
\textbf{Disposition} & \textbf{A: arithmetic} & \textbf{B: GPU parallelism} & \textbf{C: ML numerical behavior} & \textbf{D: Broadcast/\allowbreak{}layout} \\
\midrule
\endhead
\midrule
\multicolumn{5}{r@{}}{\textit{Continued on next page}}\\
\endfoot
\bottomrule
\endlastfoot
\textbf{Generated probes} & 757 & 702 & 178 & 9 \\[4pt]
\textbf{Compile/\allowbreak{}stillborn} & 56 & 3 & 45 & 1 \\[4pt]
\textbf{Host-control failure} & 196 & 172 & 70 & 3 \\[4pt]
\textbf{Other excluded} & 20 & 5 & 0 & 0 \\[4pt]
\textbf{Anchor-killed} & 186 & 337 & 13 & 3 \\[4pt]
\textbf{Harness-scoped equivalence} & 9 & 0 & 0 & 0 \\[4pt]
\textbf{Survived, later falsified} & 25 & 42 & 10 & 0 \\[4pt]
\textbf{Unresolved} & 265 & 143 & 40 & 2 \\[4pt]
\end{xltabular}
\endgroup
\par\medskip

\noindent\textbf{Host controls.} The 441 host-control exclusions arise from 22 distinct hosts: eight with state-binding failures, eight with output mismatches, three with candidate execution failures, one with an instantiation failure, and two with CUDA configuration or argument errors. The other 68 hosts pass control. Dynamic evidence uses these passing controls; one independently discharged static-equivalence probe belongs to a host without a valid dynamic control, so the retained probe population spans 69 host identities.

\paragraph{Operator catalog.}
Host identities agree with the source inventory.
The operator families contain three arithmetic, four GPU-parallelism, seven ML-numerical, and two broadcast/layout operators.
Probe records identify 15 operators with instances; \nolinkurl{reduction_reorder} has none.
Table~\ref{tab:app-operators} pairs recorded counts with target constructs and rewrites described by the available implementation.
Its exact source bytes are not bound to the C1 generation-time implementation, so the descriptions neither establish the complete matching constraints used for this batch nor guarantee its regeneration.

\Needspace{8\baselineskip}
\begingroup
\small
\setlength{\tabcolsep}{4pt}
\begin{xltabular}{\linewidth}{@{}l p{0.29\linewidth}r X@{}}
\caption{Mutation-operator counts and descriptive rewrites.}\label{tab:app-operators}\\
\toprule
\textbf{Family} & \textbf{Operator} & \textbf{Probes} & \textbf{Descriptive target and rewrite} \\
\midrule
\endfirsthead
\caption[]{Mutation-operator counts and descriptive rewrites. (continued)}\\
\toprule
\textbf{Family} & \textbf{Operator} & \textbf{Probes} & \textbf{Descriptive target and rewrite} \\
\midrule
\endhead
\midrule
\multicolumn{4}{r@{}}{\textit{Continued on next page}}\\
\endfoot
\bottomrule
\endlastfoot
A & \nolinkurl{arith_replace} & 268 & Binary arithmetic: $+\leftrightarrow-$ and $\times\leftrightarrow/$ \\
A & \nolinkurl{relop_replace} & 267 & Relational operators: strict/non-strict and equality/inequality swaps \\
A & \nolinkurl{const_perturb} & 222 & Numeric literals: integer $\pm1$; floating point $\times1.01$ or $\times0.99$ \\
B & \nolinkurl{index_replace} & 267 & Program/thread/block indices: change the selected axis \\
B & \nolinkurl{sync_remove} & 104 & Matched GPU barriers: remove the synchronization call \\
B & \nolinkurl{mask_boundary} & 219 & Boundary guards: change comparison strictness or offset the bound \\
B & \nolinkurl{launch_config_mutate} & 112 & Matched grid/block sizing expressions: add or subtract one \\
C & \nolinkurl{stab_remove} & 7 & Matched max-subtraction stabilization: remove the subtraction \\
C & \nolinkurl{acc_downgrade} & 46 & Matched accumulator or conversion expressions: lower precision \\
C & \nolinkurl{epsilon_modify} & 32 & Matched small epsilon literals: replace by zero or $10^{-2}$ \\
C & \nolinkurl{scale_modify} & 33 & Matched inverse-square-root scaling: remove or alter the scaling \\
C & \nolinkurl{cast_remove} & 38 & Explicit dtype conversions: remove the matched cast \\
C & \nolinkurl{reduction_reorder} & 0 & Matched sums: reverse the reduction input order \\
C & \nolinkurl{init_modify} & 22 & Min/max initializers: replace $\pm\infty$ with finite values \\
D & \nolinkurl{broadcast_unsafe} & 4 & Explicit expand/broadcast/unsqueeze: remove the matched operation \\
D & \nolinkurl{layout_assume} & 5 & Matched contiguity conversions or checks: remove the call \\
\midrule
\multicolumn{2}{@{}l}{\textbf{Total}} & \textbf{1,646} & \\
\end{xltabular}
\endgroup
\par\medskip

\paragraph{Static-screen checks.}
The \texttt{dead\_\allowbreak{}host\_\allowbreak{}constant} screen examines assignments outside classes and functions in parseable Python source.
For \texttt{const\_\allowbreak{}perturb} changes, it checks that the assigned name does not occur directly in the \texttt{Model}/\texttt{ModelNew} class AST.
This syntactic check alone does not establish dependency isolation: the full source difference, other uses, CUDA strings, initialization, forward computation, and model state must also be considered.
For the nine scoped cases described in \cref{sec:study}, the unchanged CUDA strings do not interpolate the altered constant.

\paragraph{Non-transitivity example.}
Using the task reference $R$, host $H$, and mutant $M$ from \cref{sec:study}, define agreement by $|a-b|\leq0.01+0.01|b|$.
For $R=1$, $H=1.019$, and $M=0.981$, both $H$ and $M$ agree with $R$, but $|M-H|=0.038>0.02019$.
This example illustrates why a host-relative witness does not imply a reference-relative failure.

\Needspace{10\baselineskip}
\subsection{Recommendation Reproducibility and Observation Coverage}
\label{sec:app-recommendations}

\paragraph{Fold and ranking details.}
The task-held-out construction in \cref{sec:study} assigns folds by hashing task identity with the fixed \texttt{e1-crossfit-v1} prefix.
Missing template identities remain missing; only explicitly recorded random trials map to IID.
For tie-breaking, mean training time is rounded to 0.1\,ms before canonical template identity is compared.
The artifact's complete 75-probe manifest identifies templates, folds, and recommendation outcomes.

\paragraph{Conclusive observations.}
The coverage partitions and recommendation counts are in \cref{tab:crossfit}.
Class-conditioned lists contain one to five recommendations (mean 3.04 per probe).
Table~\ref{tab:app-observability} supplies the additional probe--template observation counts.
Its denominators count recommended pairs; the numerators count pairs with conclusive records.
Joint-complete probes have two nonempty, fully conclusive recommendation lists.

\Needspace{8\baselineskip}
\begingroup
\small
\setlength{\tabcolsep}{4pt}
\begin{xltabular}{\linewidth}{@{}R{0.35}R{1.25}R{1.25}R{1.15}@{}}
\caption{Observation coverage of recommended probe--template pairs.}\label{tab:app-observability}\\
\toprule
\textbf{Cap} & \textbf{Class templates conclusive/total} & \textbf{Global templates conclusive/total} & \textbf{Joint-complete probes} \\
\midrule
\endfirsthead
\caption[]{Observation coverage of recommended probe--template pairs. (continued)}\\
\toprule
\textbf{Cap} & \textbf{Class templates conclusive/total} & \textbf{Global templates conclusive/total} & \textbf{Joint-complete probes} \\
\midrule
\endhead
\midrule
\multicolumn{4}{r@{}}{\textit{Continued on next page}}\\
\endfoot
\bottomrule
\endlastfoot
1 & 31/75 & 6/75 & 6 \\
2 & 100/146 & 58/150 & 5 \\
3 & 139/192 & 104/225 & 2 \\
4 & 155/216 & 127/300 & 2 \\
5 & 160/228 & 139/375 & 2 \\
6 & 160/228 & 152/450 & 1 \\
7 & 160/228 & 164/525 & 0 \\
8 & 160/228 & 178/600 & 0 \\
\end{xltabular}
\endgroup
\par\medskip

Conclusive-template proportions are 41.3\% versus 8.0\% at cap one and 70.2\% versus 29.7\% at cap eight.
Neither list witnesses a failure among the joint-complete probes at any cap; these cohorts shrink from six to zero and do not support an equivalence conclusion.
For the common observable-pool analysis in \cref{sec:rq4}, the total recommendations per ranking are 73, 130, and 136 at $k=1,2,3$.
At $k\geq4$, both exhaust the pools and select the same 138 probe--template pairs.

\paragraph{Deployed-prior traceability.}
The recorded unsupported association is \nolinkurl{L1_P37__index_replace__11}, whose stored \texttt{F-IDX} support for \texttt{boundary\_last\_element} is 1/1; its admissibility record rejects the last-element witness for kernel nondeterminism at that position.
This explanation reports the recorded reason, not a new nondeterminism experiment.
Confirmed \texttt{F-IDX} evidence contains 156 probes, with head-heavy support 6/6 and IID support 150/156, but no admissible last-element witness.
These support statistics do not replace the deployed prior.

All 9,482 selected directed observations match the saved plans by original test identity, policy, mode, parameters, seed, and candidate/task.
The two last-element cases per directed plan give 1,676 planned cases.
In addition to the full-population masking results in \cref{sec:rq4}, the B1u-accepted cohort retains 496/498 discoveries and 118/120 native discoveries.
The affected C5 L1-P37 source IDs are \nolinkurl{14e554b1d46601b9} and \nolinkurl{67d428bcecfba425}.
Each has one FP32 output element violating tolerance at shape $[112,64,512,512]$, with reported maximum absolute differences $1.35496\times10^{-4}$ and $1.35243\times10^{-4}$, respectively.
These remain unadjudicated numerical witnesses.
The schedules do not identify each case's generating association; order, seeds, stopping, discovery, and cost under a regenerated plan remain unmeasured.

\Needspace{10\baselineskip}
\subsection{Witness Classification and Traceability}
\label{sec:app-witnesses}

The named-stress witnesses counted in \cref{sec:rq4} comprise 62 screening and seven enhanced-search records.
A screening witness requires the label \texttt{WITNESSED\_\allowbreak{}NON\_\allowbreak{}EQUIVALENT}.
Enhanced-search witnesses require \texttt{FALSIFIED} in the probe's terminal record, selected by its latest recorded completion time.
The two FP16 observations have matched passing host controls and join through the recorded subject-to-probe map.

The case classifier reads \nolinkurl{divergence.round_type} and \nolinkurl{divergence.policy}: an explicit stress round with a named policy is a stress witness, and an explicit random round is a random witness.
FP16 cases use the directly recorded dtype.
A reproduced-crash record without a template retains that unknown identity.
The artifact binds all 77 probes to source-file and original-line hashes, anchor records, and available matching trials.

For example, \nolinkurl{L2_P81__index_replace__3} has \texttt{round\_type=random}, a null policy, and seed 10000.
\nolinkurl{L1_P95__relop_replace__5} records a reproduced crash without a divergence template, so no IID or stress policy is imputed.

\Needspace{10\baselineskip}
\subsection{Additional Dtype Evidence}
\label{sec:app-dtype}

Table~\ref{tab:app-dtype-shapes} records the input and output shapes for the two cases summarized in \cref{tab:dtype-cases}.

\Needspace{8\baselineskip}
\begingroup
\small
\setlength{\tabcolsep}{4pt}
\begin{xltabular}{\linewidth}{@{}p{0.18\linewidth}X@{}}
\caption{Operation and shape details for the dtype cases in Table~\ref{tab:dtype-cases}.}\label{tab:app-dtype-shapes}\\
\toprule
\textbf{Candidate} & \textbf{Operation/network and input-to-output shapes} \\
\midrule
\endfirsthead
\caption[]{Operation and shape details for the dtype cases in Table~\ref{tab:dtype-cases}. (continued)}\\
\toprule
\textbf{Candidate} & \textbf{Operation/network and input-to-output shapes} \\
\midrule
\endhead
\midrule
\multicolumn{2}{r@{}}{\textit{Continued on next page}}\\
\endfoot
\bottomrule
\endlastfoot
C3-23530 & LeNet-5 with four ReLU replacements; \texttt{[1,\allowbreak{}1,\allowbreak{}32,\allowbreak{}32]} $\to$ \texttt{[1,\allowbreak{}10]} logits \\[4pt]
C3-00995 & \texttt{torch.\allowbreak{}matmul(A,\allowbreak{}B)}, \texttt{A:[16,\allowbreak{}1024,\allowbreak{}2048]}, \texttt{B:[2048,\allowbreak{}768]}; boundary \texttt{[16,\allowbreak{}1024,\allowbreak{}768]}, readout \texttt{[16,\allowbreak{}8]} \\[4pt]
\end{xltabular}
\endgroup
\par\medskip

For C3-00995, the FP16 custom path returns FP16 outputs.
An additional case, C5-01707, uses LeNet with a custom ReLU and input shape $[1,1,32,32]$, producing $[1,10]$ logits.
All 48 applicable FP32 inputs agree with custom execution established.
FP16 and BF16 fail at the custom Float pointer requirement.
This case indicates an additional-dtype portability limitation, separate from the recovered FP32 evaluation setting.

\Needspace{10\baselineskip}
\subsection{Replacement Outcome Details}
\label{sec:app-replacement}

Table~\ref{tab:app-other-outcomes} expands the Other category in \cref{tab:replacement}.
Codes preserve the recorded configuration classification, with each configuration contributing to exactly one row.
Adapter/reference limitations and lack of evidence for custom execution remain separate from candidate computation failures.

\Needspace{34\baselineskip}
\begingroup
\small
\setlength{\tabcolsep}{4pt}
\begin{xltabular}{\linewidth}{@{}l X r@{}}
\caption{The 19 recorded terminal categories within the 523 Other outcomes.}\label{tab:app-other-outcomes}\\
\toprule
\textbf{Code} & \textbf{Terminal outcome} & \textbf{Configurations} \\
\midrule
\endfirsthead
\caption[]{The 19 recorded terminal categories within the 523 Other outcomes. (continued)}\\
\toprule
\textbf{Code} & \textbf{Terminal outcome} & \textbf{Configurations} \\
\midrule
\endhead
\midrule
\multicolumn{3}{r@{}}{\textit{Continued on next page}}\\
\endfoot
\bottomrule
\endlastfoot
\texttt{BUILD} & Extension build failure & 174 \\
\texttt{CUPY} & CuPy dependency or execution path unavailable & 12 \\
\texttt{SO} & Shared-library loading failure & 3 \\
\texttt{SYNTAX} & Candidate source syntax failure & 3 \\
\texttt{UNDEFINED} & Undefined name during candidate initialization & 9 \\
\texttt{CONSTRUCTOR} & Candidate constructor argument mismatch & 36 \\
\texttt{FORWARD} & Adapter-generated call argument mismatch & 87 \\
\texttt{CSTATE} & Candidate state mapping unavailable & 81 \\
\texttt{RSTATE} & Reference state mapping unavailable & 81 \\
\texttt{NO\_ENTRY} & No callable binding to the custom entry & 4 \\
\texttt{PLACEHOLDER} & Placeholder implementation & 3 \\
\texttt{DISABLED} & Custom computation disabled by the original configuration & 9 \\
\texttt{ATEN\_ONLY} & ATen-only execution & 1 \\
\texttt{CUBLAS} & Library delegation without handwritten-kernel execution evidence & 1 \\
\texttt{REF\_DTYPE} & Reference dtype incompatibility & 14 \\
\texttt{REF\_NONFINITE} & Nonfinite reference output & 1 \\
\texttt{LAUNCH} & Custom-path launch failure & 1 \\
\texttt{PROCESS} & Process termination without a valid result & 1 \\
\texttt{TIMEOUT} & Timeout before planned completion & 2 \\
\midrule
\multicolumn{2}{@{}l}{\textbf{Total}} & \textbf{523} \\
\end{xltabular}
\endgroup
\par\medskip

The five direct-comparison cases without an applicable downstream readout are C2-00096, C3-10177, C3-10342, C3-10422, and C3-11178.
These cases do not establish downstream agreement.
Table~\ref{tab:app-replacement-shapes} supplies the input and reference-output shapes for the four witnesses in \cref{tab:replacement-cases}.

\Needspace{8\baselineskip}
\begingroup
\small
\setlength{\tabcolsep}{4pt}
\begin{xltabular}{\linewidth}{@{}p{0.18\linewidth}X@{}}
\caption{Input and expected-output shapes for the witnesses in Table~\ref{tab:replacement-cases}.}\label{tab:app-replacement-shapes}\\
\toprule
\textbf{Candidate} & \textbf{Input $\to$ expected output} \\
\midrule
\endfirsthead
\caption[]{Input and expected-output shapes for the witnesses in Table~\ref{tab:replacement-cases}. (continued)}\\
\toprule
\textbf{Candidate} & \textbf{Input $\to$ expected output} \\
\midrule
\endhead
\midrule
\multicolumn{2}{r@{}}{\textit{Continued on next page}}\\
\endfoot
\bottomrule
\endlastfoot
C3-17320 & \texttt{[128,\allowbreak{}10]} $\to$ scalar \texttt{[]} \\[4pt]
C3-20564 & \texttt{[128,\allowbreak{}512]} $\to$ \texttt{[1,\allowbreak{}8]} \\[4pt]
C5-00631 & \texttt{[16,\allowbreak{}64,\allowbreak{}256,\allowbreak{}256]} $\to$ \texttt{[16,\allowbreak{}8]} \\[4pt]
C5-01485 & \texttt{[10,\allowbreak{}3,\allowbreak{}224,\allowbreak{}224]} $\to$ \texttt{[10,\allowbreak{}1000]} \\[4pt]
\end{xltabular}
\endgroup
\par\medskip

For C3-20564, the boundary comparison has shapes $[1,256,128,1]$ versus $[128,1]$.
For C5-00631, the failure at input 37 follows 36 passing inputs.
For C5-01485, the published \texttt{forward()} invokes \texttt{self.model} followed by the classifier.
The native reference block is PyTorch Conv2d + BatchNorm2d + ReLU.
The replacement invokes the published \texttt{conv\_\allowbreak{}bn\_\allowbreak{}relu\_\allowbreak{}cuda} and then the remaining MobileNetV1 layers.
The subsequent layer reports \texttt{expected 4D input (got 3D input)}.
These entry bindings complement the custom-execution evidence in \cref{sec:replacement}.

\end{document}